\documentclass[aps,prd,english,preprintnumbers,nofootinbib,floatfix,twocolumn,10pt]{revtex4}

\usepackage{amsfonts,amsmath,amssymb}
\usepackage{graphicx,epsfig}
\usepackage{float}
\usepackage{subfig}
\usepackage{subfloat}
\usepackage[utf8]{inputenc}
\usepackage{hyperref}
\usepackage{babel}

\def	\beq	{\begin{equation}}
\def	\eeq	{\end{equation}}

\begin{document}

\title{Equilibrium to Einstein: Entanglement, Thermodynamics, and Gravity}

\author{Andrew Svesko}

\affiliation{Department of Physics and Beyond: Center for Fundamental Concepts in Science\\
Arizona State University, Tempe, Arizona 85287, USA\\}

\begin{abstract}
\begin{center}
{\bf Abstract}
\end{center}
\noindent
Here we develop the connection between thermodynamics, entanglement, and gravity. By attributing thermodynamics to timeslices of a causal diamond, we show that the Clausius relation $T\Delta S_{\text{rev}}=Q$, where $\Delta S_{\text{rev}}$ is the \emph{reversible} entropy change, gives rise to the non-linear gravitational equations of motion for a wide class of diffeomorphism invariant theories. We then compare the Clausius relation to the first law of causal diamond mechanics (FLCD), a geometric identity and necessary ingredient in deriving Jacobson's entanglement equilibrium proposal -- the entanglement entropy of a spherical region with a fixed volume is maximal in vacuum. Specifically we show that the condition of fixed volume can be understood as subtracting the irreversible contribution to the thermodynamic entropy. This provides a ``reversible thermodynamic process" interpretation of the FLCD, and that the condition of entanglement equilibrium may be regarded as equilibrium thermodynamics for which the Clausius relation holds. Finally, we extend the entanglement equilibrium proposal to the timelike stretched horizons of future lightcones, providing an entanglement interpretation of stretched lightcone thermodynamics.
\end{abstract}

\thispagestyle{empty}

\maketitle

\section{Overview} \label{sec:overview}
\noindent The discovery that black holes carry entropy \cite{Bekenstein72-1,Hawking74-1}, 
\beq S_{BH}=\frac{A_{\mathcal{H}}}{4G}\; , \label{BHent}\eeq
provides the two following realizations: (i) A world with gravity is holographic \cite{Susskind:1994vu}, and (ii) spacetime is emergent \cite{Jacobson95-1}. The former of these comes from the observation that the thermodynamic entropy of a black hole (\ref{BHent}) goes as the area of its horizon $A_{\mathcal{H}}$, and the latter from noting that black holes are spacetime solutions to Einstein's equations. In fact, black holes are not the only spacetime solutions which carry entropy; any solution which has a horizon, e.g., Rindler space and the de Sitter universe, also possess a thermodynamic entropy proportional to the area of their respective horizons. The fact that Rindler space carries an entropy is particularly striking as there the notion of horizon is observer dependent. This leads to the proposal that an arbitrary spacetime -- which may appear locally as Rindler space -- is equipped with an entropy proportional to the area of a local Rindler horizon, and that thermodynamic relationships, e.g., the Clausius relation $T\Delta S=Q$, have geometric meaning. Specifically, 
\beq T\Delta S=Q\Rightarrow G_{\mu\nu}+\Lambda g_{\mu\nu}=8\pi GT_{\mu\nu}\; . \label{ClausiusGrav}\eeq
That is, Einstein gravity arises from the thermodynamics of spacetime \cite{Jacobson95-1}. 

Recently it was shown how to generalize (\ref{ClausiusGrav}) to higher derivative theories of gravity \cite{Parikh:2017aas}. By attributing a temperature and entropy to a stretched future lightcone -- a timelike hypersurface composed of the worldlines of constant and uniformly radially accelerating observers -- the equations of motion for a broad class of higher derivative theories of gravity are a consequence of the Clausius relation $T\Delta S_{\text{rev}} =Q$, where $\Delta S_{\text{rev}}$ is the \emph{reversible} entropy, i.e., the entropy growth solely due to a flux of matter crossing the horizon of the stretched lightcone. This result shows that arbitrary theories of gravity arise from the thermodynamics of some underlying microscopic theory of spacetime.

Despite some successes in deriving (\ref{BHent}) in specific cases \cite{Strominger96-1,Rovelli96-1}, it is still unclear what the physical degrees of freedom encoded in $S_{BH}$ correspond to microscopically. Similarly, the underlying microscopics of spacetime giving rise to Einstein's equations is obscure. A potential explanation comes from studying entanglement entropy (EE) of quantum fields outside of the horizon. For a generic $(d+1)$ quantum field theory (QFT) with $d>1$, the EE of a region $A$ admits an area law \cite{Bombelli:1986rw,Srednicki:1993im}
\beq S^{EE}_{A}=c_{0}\frac{\mathcal{A}(\partial A)}{\epsilon^{d-1}}+\text{subleading divergences}+S_{\text{finite}}\; , \label{SQFT}\eeq 
where $\epsilon$ is a cutoff for the theory, illustrating that the EE is in general UV divergent, and $\mathcal{A}$ is the area of the $(d-1)$ boundary region $\partial A$ separating region $A$ from it's complement.  Identifying $\frac{c_{0}}{\epsilon^{d-1}}\to\frac{1}{4G}$ suggests $S_{BH}$ to be interpreted as the leading UV divergence in the EE for quantum fields outside of a horizon.

Further progress can be made when we consider quantum field theories with holographic duals. Specifically, in the context of $\text{AdS}_{d+2}/\text{CFT}_{d+1}$ duality \cite{Maldacena98-1}, one is led to the Ryu-Takayanagi (RT) conjecture \cite{Ryu06-1}:
\beq S^{EE}_{A}=\frac{\mathcal{A}(\gamma_{A})}{4G^{(d+2)}}\; , \label{RT}\eeq
which relates the EE of holographic CFTs (HEE) to the area of a $d$-dimensional (static) minimal surface $\gamma_{A}$  in $\text{AdS}_{d+2}$ whose boundary is homologous to $\partial A$.\footnote{The RT conjecture has a covariant generalization, in which the static minimal surface $\gamma_{A}$ is replaced by an extremal surface $\Sigma_{A}$, \cite{Hubeny:2007xt}.} The RT formula (\ref{RT}) is specific to CFTs dual to general relativity, and does not include quantum corrections. The proposal was proved in \cite{Lewkowycz:2013nqa}, and has been extended to include quantum corrections \cite{Faulkner13-1}, and for CFTs dual to higher derivative theories of gravity \cite{Dong:2013qoa}. When the minimal surface $\gamma_{A}$ is the horizon of a black hole, one observes that black hole entropy is equivalent to HEE, $S_{HEE}|_{\gamma_{A}=\mathcal{H}}=S_{BH}$ \cite{Casini:2011kv}. 

Similar to the situation with black hole thermodynamics, this observation suggests that gravity emerges from quantum entanglement, i.e., spacetime is built from entanglement \cite{VanRaamsdonk10-1,Bianchi12-1}. To take on this proposal, one can study the properties of HEE and look for the resulting geometric consequences. Indeed, the EE of a QFT generically satisfies a first law reminiscient of the first law of thermodynamics \cite{Blanco:2013joa,Wong:2013gua}
\beq \delta S^{EE}_{A}=\delta \langle H_{A}\rangle\; .\label{firstlawEE}\eeq
Here $\delta S^{EE}_{A}$ is the variation of the EE of region $A$, while $\delta \langle H_{A}\rangle$ is the variation of the modular Hamiltonian $H_{A}$ defined by $\rho_{A}\equiv e^{-H_{A}}$. When one specializes to the case where the region $A$ is a ball of radius $R$, the modular Hamiltonian can be identified with the thermal energy of the region. 

For holographic CFTs the first law of entanglement entropy (\ref{firstlawEE}) can be understood as a geometric constraint on the dual gravity side. By substituting (\ref{RT}) into the left hand side (LHS) of (\ref{firstlawEE}), and relating the energy-momentum tensor of the CFT to a metric perturbation in AdS, one arrives at the \emph{linearized} Einstein equations \cite{Lashkari13-1}:
\beq \delta S^{EE}_{A}=\delta \langle H_{A}\rangle \Rightarrow G_{\mu\nu}+\Lambda g_{\mu\nu}=8\pi GT_{\mu\nu}\; .\eeq
By considering the higher derivative gravity generalization of (\ref{RT}), similar arguments lead to the linearized equations of motion for higher derivative theories of gravity \cite{Faulkner13-2}. The non-linear behavior of gravitational equations of motion is encoded in a generalized form of (\ref{firstlawEE}), where one must take into account the relative entropy of excited CFT states \cite{Faulkner:2017tkh,Haehl:2017sot}. In this way, gravity emerges from spacetime entanglement.

Recently it has been shown how to derive gravitational equations of motion from entanglement considerations without explicit reference to AdS/CFT duality, and is therefore slightly more general than the derivation in \cite{Lashkari13-1,Faulkner13-2}. This approach, first proposed by Jacobson, is the \emph{entanglement equilibrium} conjecture \cite{Jacobson16-1}, which can be stated as follows: \emph{In a theory of quantum gravity, the entanglement entropy of a spherical region with a fixed volume is maximal in the vacuum}. This hypothesis relies on assuming that the quantum theory of gravity is UV finite (as is the case in string theory) and therefore yields a finite EE, where the cutoff $\epsilon$ introduced in (\ref{SQFT}) is near the Planck scale, $\epsilon\sim\ell_{P}$, and being able to identify the entanglement entropy $S_{EE}^{A}$ with the generalized entropy $S_{\text{gen}}$, which is independent of $\epsilon$ \cite{Susskind:1994sm,Solodukhin:2011gn}:
\beq S_{EE}^{A}=S_{\text{gen}}=S^{(\epsilon)}_{BH}+S^{(\epsilon)}_{\text{mat}}\;.\label{Sgen}\eeq
Here $S^{(\epsilon)}_{BH}$ is the Bekenstein-Hawking entropy (\ref{BHent}) expressed in terms of renormalized gravitational couplings, and $S^{(\epsilon)}_{\text{mat}}$ is the renormalized EE of matter fields. The generalized entropy $S_{\text{gen}}$ is independent of $\epsilon$ as the renormalization of gravitational couplings is achieved via the matter loop divergences. 

When one interprets the EE as the generalized entropy, one may therefore assign EE to surfaces other than cross sections of black hole horizons, or the minimal surfaces identified in the RT formula (\ref{RT}). In this way, without assuming holographic duality, one discovers a connection between geometry and entanglement entropy. Furthermore, taking into consideration the underlying thermodynamics of spacetime \cite{Jacobson95-1}, this link provides a route to derive dynamical equations of gravity -- not from thermodynamics, but from entanglement. 

With these consderations in mind, the variation of the EE of a spherical region at fixed volume is given by 
\beq \delta S^{A}_{EE}|_{V}=\frac{\delta A|_{V}}{4G}+\delta S_{\text{mat}}=0\;,\eeq
i.e., the vacuum is in a maximal entropy state. In the case of small spheres, this entanglement equilibrium condition is equivalent to imposing the full non-linear Einstein equations at the center of the ball \cite{Jacobson16-1}. Recently this maximal entropy condition has been generalized to include higher derivative theories of gravity, where $S^{(\epsilon)}_{BH}$ in (\ref{Sgen}) is replaced by the higher derivative extension of gravitational entropy, the Wald entropy $S^{(\epsilon)}_{\text{Wald}}$, in which case the maximal entropy condition becomes
\beq \delta S^{A}_{EE}|_{W}=\delta S_{\text{Wald}}|_{W}+\delta S_{\text{mat}}=0\;,\eeq
where the volume $V$ must be replaced with a new local geometrical quantity called the \emph{generalized volume} $W$. This condition, when applied to small spheres, is equivalent to imposing the linearized equations of motion for a higher derivative theory of gravity \cite{Bueno16-1}. 

Here we aim to extend the work of \cite{Parikh:2017aas} and \cite{Bueno16-1} and develop the connection between thermodynamics, entanglement, and gravity. Specifically, we will consider the geometric set-up of \cite{Bueno16-1} and provide a ``physical process" derivation of the geometric identity known as the \emph{first law of causal diamond mechanics} (FLCD) crucial in deriving the entanglement equilibrium condition. We accomplish this as follows: First attribute thermodynamics to sections of causal diamonds in an arbitrary spacetime, and compute the Clausius relation $T\Delta S_{\text{rev}}=Q$, where $\Delta S_{\text{rev}}$ is the \emph{reversible} (gravitational) entropy change computed via a reversible thermodynamic process. Using the techniques developed in \cite{Parikh:2017aas}, we will then show that the Clausius relation is geometrically equivalent to the non-linear gravitational equations of motion for a broad class of diffeomorphism invariant theories of gravity, thereby connecting gravity to thermodynamics. Next, we show how the FLCD relates to the Clausius condition, by explicitly showing that the leading contribution to generalized volume $\bar{W}$ is precisely the entropy change due to the natural increase of the causal diamond, presenting an equivalence between entanglement equilibrium and (reversible) equilibrium thermodynamics in theories of gravity. 

Then, noting the geometric similarities of causal diamonds and stretched lightcones, we will derive a ``first law of stretched lightcones", and show that it is equivalent to an entanglement equilibrium condition. This not only sheds light on the microscopic origins of the thermodynamics of stretched lightcones, but also provides another derivation of the non-linear (semi-classical) Einstein equations and (linearized) equations of motion of higher derivative theories of gravity from spacetime entanglement. We will also discuss why the Clausius relation gives rise to the non-linear equations while the entanglement equilibrium condition gives only the linearized equations for higher derivative theories of gravity. This will help us better understand  how both \cite{Bueno16-1} and the work here may be extended to include the non-linear contributions of higher derivative equations of motion.

The outline of the paper is as follows: We begin by reviewing the geometric set-up of the stretched lightcone and causal diamond in section (\ref{sec:geosetup}) and observe the similarities between the two constructions. In section (\ref{sec:thermoCD}) we present an alternative derivation of the FLCD and show how it relates to the Clausius relation $T\Delta S_{rev} =Q$ applied to the diamond. We further show that the Clausius relation is geometrically equivalent to the full non-linear gravitational equations of motion for a broad class of diffeomorphism invariant theories of gravity. A first law of stretched lightcones is developed in section (\ref{sec:entLC}), where we show that it is equivalent to an entanglement equilibrium condition, which we also illustrate is equivalent to the \emph{linearized} gravitational equations of motion being satisfied. 


\section{Geometry of Stretched Lightcones and Causal Diamonds}
 \label{sec:geosetup}
\indent

\subsection{Stretched Lightcones}
\indent

We begin with a review of the construction of the stretched lightcone (for more details see \cite{Parikh:2017aas}). For concreteness, let us first restrict to pure $D$-dimensional Minkowski space. In Minkowski space there are $\binom{D+1}{2}$ independent Killing vectors $\chi^{a}$ corresponding to spacetime translations and Lorentz transformations. The flow lines of Cartesian boost vectors, e.g., $x\partial^{a}_{t}+t\partial^{a}_{x}$, trace the worldlines of Rindler observers, i.e., observers traveling with constant acceleration in some Cartesian direction. 

The stretched future lightcone can be viewed as a spherical Rindler horizon generated by the radial boost vector:
\beq \xi^{a}\equiv r\partial^{a}_{\;t}+t\partial^{a}_{\;r}=\sqrt{x^{i}x_{i}}\partial^{a}_{t}+\frac{tx^{j}}{\sqrt{x^{i}x_{i}}}\partial^{a}_{\;j}\;, \label{xi}\eeq
where $r$ is the radial coordinate and $x^{i}$ are spatial Cartesian coordinates. We define the stretched future lightcone as a congruence of worldlines generated by these radial boosts. Unlike their Cartesian boost counter-parts, which preserve local Lorentz symmetry, the radial boost vector is not a Killing vector in Minkowski space; this is because radial boosts are not isometries in Minkowski space. 

The flow lines of $\xi^{a}$ trace out hyperbolae in Minkowski space. Let us define a codimension-1 timelike hyperboloid via the set of curves which obey
\beq r^{2}_{\text{Mink}}-t^{2}=\alpha^{2}\;,\eeq
where $t\geq0$ and $\alpha$ is some length scale with dimensions of length. This hyperboloid can be understood as a stretched future lightcone emanating from a point $p$ at the origin. The constant-$t$ sections of the hyperboloid are $(D-2)$-spheres with an area given by 
\beq A_{\text{Mink}}(t)=\Omega_{D-2}(\alpha^{2}+t^{2})^{(D-2)/2}\eeq
Here we have that $\xi^{2}=-\alpha^{2}$, and is therefore an unnormalized tangent vector to the worldlines of the spherical Rindler observers. The normalized velocity vector is defined as $u^{a}=\xi^{a}/\alpha$, with $u^{2}=-1$, and has a proper acceleration with magnitude
\beq a_{\text{Mink}}=\frac{1}{\alpha}\;.\eeq
The stretched future lightcone, in Minkowski space, can therefore be understood as a congruence of worldlines of a set of constant radially accelerating observers, all with the same uniform acceleration of $1/\alpha$. 

Let us now consider what happens in an arbitrary spacetime. In the vicinity of any point $p$, spacetime is locally flat. The components of a generic metric tensor can always be expanded using Riemann normal coordinates (RNC):
\beq g_{ab}(x)=\eta_{ab}-\frac{1}{3}R_{acbd}(p)x^{c}x^{d}+...\;, \label{RNC}\eeq
where the Riemann tensor is evaluated at the point $p$, the origin of the RNC system. Here $x^{a}$ are Cartesian coordinates and $\eta_{ab}$ is the Minkowski metric in Cartesian coordinates. Since a generic spacetime is locally flat, there still exist the $\binom{D+1}{2}$ vectors $\chi^{a}$ which preserve the isometries of Minkowski space, locally, however, they are no longer exact Killing vectors; the presence of quadratic terms $\mathcal{O}(x^{2})$ in the RNC expansion (\ref{RNC}) indicates that these vectors will not satisfy Killing's equation and Killing's identity at some order in $x$. The specific order depends on the nature of the vector $\chi^{a}$, e.g., for Lorentz boosts the components are of order $\mathcal{O}(x)$. Therefore, for the generators of local Lorentz transformations, Killing's equation and Killing's identity will fail as
\beq \nabla_{a}\chi_{b}+\nabla_{b}\chi_{a}\approx\mathcal{O}(x^{2})\;,\quad \nabla_{a}\nabla_{b}\chi_{c}-R^{d}_{\;abc}\chi_{d}\approx\mathcal{O}(x)\;.\eeq
We call these local Cartesian boost vectors $\chi^{a}$ approximate Killing vectors. 

The radial boost vector (\ref{xi}) is therefore not a Killing vector in an arbitrary spacetime for two reasons: (i) It is not a Killing vector in Minkowski space, and (ii) the addition of curvature via the RNC expansion leads to a further failure of Killing's equation and Killing's identity. Specifically, 
\beq
\begin{split}
&\nabla_{t}\xi_{t}=0+\mathcal{O}(x^{2})\;,\quad \nabla_{t}\xi_{i}+\nabla_{i}\xi_{t}=0+\mathcal{O}(x^{2})\;,\\
&\nabla_{i}\xi_{j}+\nabla_{j}\xi_{i}=\frac{2t}{r}\left(\delta_{ij}-\frac{x_{i}x_{j}}{r^{2}}\right)+\mathcal{O}(x^{2})\;.
\end{split}
\eeq
Observe that the $t-t$ and $t-i$ components satisfy Killing's equation at $\mathcal{O}(1)$, while the $i-j$ components fail to obey Killing's equations even at leading order. This means that Killing's identity will also fail; in fact it fails to order $\mathcal{O}(x^{-1})$. We also note that on the $t=0$ surface our radial boost vector is an instantaneous Killing vector. 

\begin{figure}[H]
\centering
 \includegraphics[width=9.3cm]{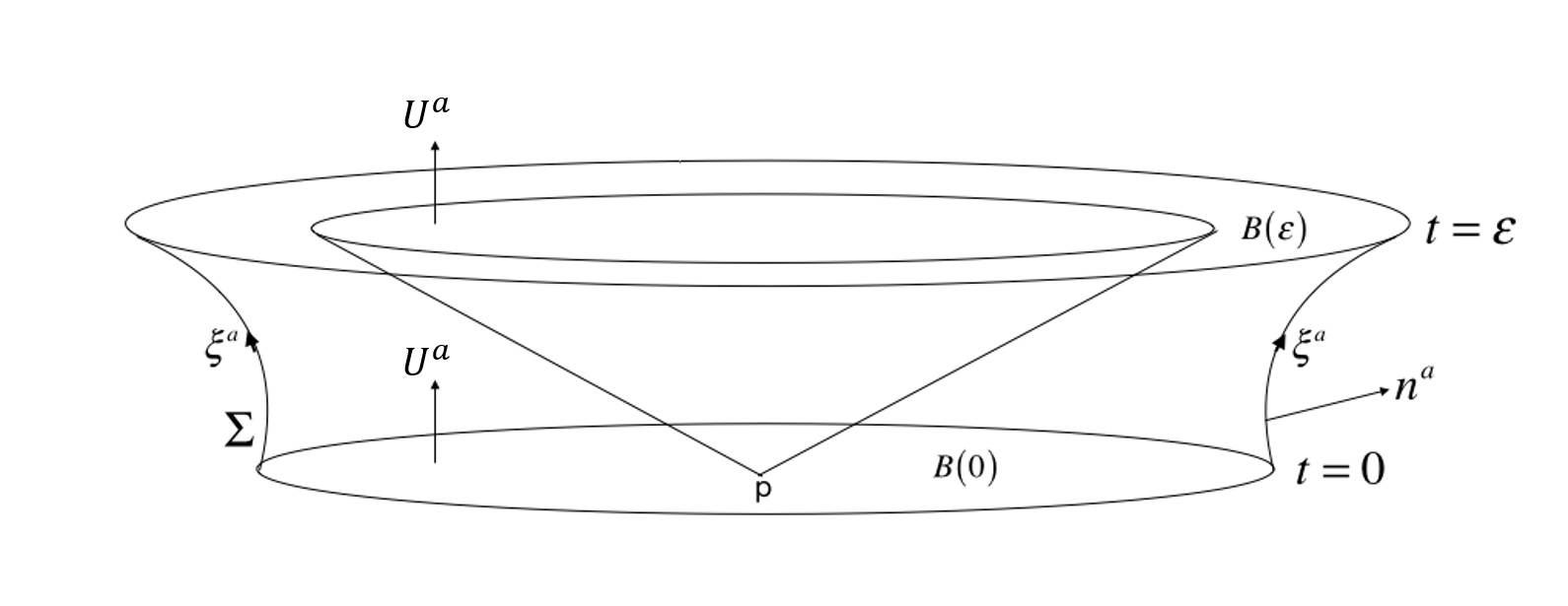}
 \caption{A congruence of radially accelerating worldlines $\xi^{a}$ with the same uniform proper acceleration $1/\alpha$ generates the stretched future light cone of point $p$, and describes a timelike hypersurface, $\Sigma$, with unit outward-pointing normal $n^{a}$. The boundary of $\Sigma$ consists of the two codimension-two surfaces $\partial \Sigma(0)$ and $\partial \Sigma(\epsilon)$ given by the constant-time slices of $\Sigma$ at $t=0$ and $t=\epsilon$, respectively. The co-dimension-1 spatial ball $B$ is the filled in co-dimension-2 surface $\partial \Sigma$.
}
 \label{sigma}\end{figure}

In an arbitrary spacetime our notion of stretched future lightcone must be modified. In a curved spacetime it is straightforward to show that 
\beq \xi^{2}=-\alpha^{2}+\mathcal{O}(x^{4})\quad a=\frac{1}{\alpha}\left(1+\mathcal{O}(x^{4})\right)\;.\eeq
Motivated by the stretched horizon defined in the black hole membrane paradigm \cite{Price86-1}, we define the stretched future lightcone $\Sigma$ as follows: Pick a small length scale\footnote{``Small" here means $\alpha$ is much smaller than the smallest curvature scale at the point $p$, i.e., the metric is taken to be roughly flat to a coordinate distance $\alpha$ from the origin.}. Then select a subset of observers who at time $t=0$ have a proper acceleration $1/\alpha$. If we follow the worldlines of these observers we would find that generically they would not have the same proper acceleration at a later generic time. This problem can be remedied by choosing a timescale $\epsilon\ll\alpha$. Over this timescale the initially accelerating observers have an approximate constant proper acceleration, and the stretched future lightcone $\Sigma$ can be regarded as a worldtube of a congruence of observers with the same nearly-constant approximately outward radial acceleration $1/\alpha$, as can be seen in figure (\ref{sigma}). With this definition, therefore, $\Sigma$ can be interpreted as a surface with constant Unruh-Davies temperature $T\equiv a/2\pi$. 

\subsection{Causal Diamonds}
\indent

\begin{figure}[H]
\includegraphics[width=6.5cm]{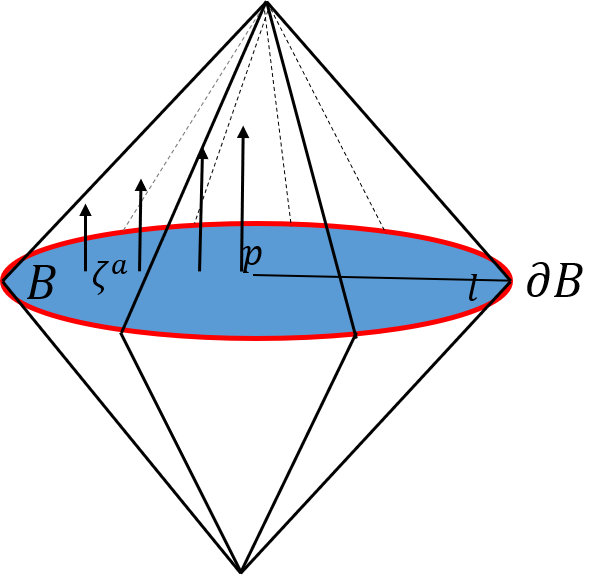}
\centering
\caption{The causal diamond as the union of future and past domains of dependence of the spatial balls $B$ of size $\ell$ with boundary $\partial B$. The diamond admits a conformal Killing vector $\zeta^{a}$ whose flow preserves the diamond, and vanishes at the boundary $r=\pm\ell$.}
\label{causaldiamond2}\end{figure}

In a maximally symmetric background, a causal diamond can be defined as the union of future and past domains of dependence of its spatial slices, balls $B$ of size $\ell$ with boundary $\partial B$. The diamond admits a conformal Killing vector (CKV) $\zeta^{a}$ whose flow preserves the diamond (see figure (\ref{causaldiamond2})). 

Conformal Killing vectors are those which satisfy conformal Killing's equation 
\beq \nabla_{a}\zeta_{b}+\nabla_{b}\zeta_{a}=2\Omega g_{ab}\;,\label{confKeqn}\eeq
where $\Omega$ satisfies
\beq \Omega=\frac{1}{D}\nabla_{c}\zeta^{c}\;,\eeq
and is related to the conformal factor $\omega^{2}$ of $\bar{g}_{ab}=\omega^{2}g_{ab}$ via $2\Omega=\zeta^{c}\nabla_{c}\ln\omega^{2}$. 

Conformal Killing vectors also satisfy the conformal Killing identity
\beq \nabla_{b}\nabla_{c}\zeta_{d}=R^{e}_{\;bcd}\zeta_{e}+(\nabla_{c}\Omega)g_{bd}+(\nabla_{b}\Omega)g_{cd}-(\nabla_{d}\Omega)g_{bc}\;.\label{confKid}\eeq
Following the discussion above, in an arbitrary spacetime the conformal Killing vectors will become approximate conformal Killing vectors, failing to satisfy the conformal Killing equation to order $\mathcal{O}(x^{2})$ in a RNC expansion about some point $p$, and the conformal Killing identity to $\mathcal{O}(x)$. 

We can define a timelike normal $U^{a}$ to $B$ via
\beq U_{a}=N\nabla_{a}\Omega\;,\eeq
with
\beq N=||\nabla_{a}\Omega||^{-1}\;,\eeq
being some normalization such that $U^{2}=-1$. In fact, it can be shown in general that
\beq N=\frac{D-2}{\kappa K}\;, \label{NKk}\eeq
where $\kappa$ is the surface gravity and $K$ is the trace of the extrinsic curvature.

One also has
\beq \nabla_{d}(\mathcal{L}_{\zeta}g_{ab})|_{B}=\frac{2}{N}U_{d}g_{ab}\quad \nabla_{a}\zeta_{b}|_{\partial B}=\kappa N_{ab}\;,\eeq
where we have the binormal $N_{ab}=2U_{[a}N_{b]}$, where $N_{a}$ is the spacelike unit normal to $U_{b}$. The spatial slice $B$ is taken to be the $t=0$ slice. 

For concreteness, in $D$-dimensional Minkowski space, the CKV which preserves the causal diamond is \cite{Bueno16-1}
\beq
\begin{split}
 \zeta^{a}&=\left(\frac{\ell^{2}-r^{2}-t^{2}}{\ell^{2}}\right)\partial^{a}_{t}-\frac{2rt}{\ell^{2}}\partial^{a}_{r}\\
& = \left(\frac{\ell^{2}-r^{2}-t^{2}}{\ell^{2}}\right)\partial^{a}_{t}-\frac{2x^{i}t}{\ell^{2}}\partial^{a}_{i}\;.\label{zeta}
\end{split}
\eeq
We point out that $\zeta^{a}$ goes null on the boundary, $t=\ell\pm r$, and $\zeta^{2}=-1$ when $r=t=0$. We also have
\beq U^{a}=\partial^{a}_{t}\quad N^{a}=\partial^{a}_{r}\Rightarrow N_{ab}=2\nabla_{[a}r\nabla_{b]}t\;,\eeq
\beq \Omega=-\frac{2t}{\ell^{2}}\quad \nabla_{a}\Omega=-\frac{2\nabla_{a}t}{\ell^{2}}=2\frac{U_{a}}{\ell^{2}}\;,\eeq
and, 
\beq N=\frac{\ell^{2}}{2}\quad K_{\partial B}=\frac{(D-2)}{\ell}\;. \label{NandK}\eeq
We see that the causal diamond has constant extrinsic curvature, constant surface gravity $\kappa=2/\ell$, and $\zeta^{a}$ is an exact Killing vector on the $t=0$ surface $B$. 

Let us remark on the similarities between the radial boost vector $\xi_{a}$ (\ref{xi}) generating the stretched future lightcone, and the conformal Killing vector $\zeta_{a}$ (\ref{zeta}) preserving the causal diamond. Specifically, we find that $\xi_{a}$ satisfies
\beq \nabla_{a}\xi_{b}+\nabla_{b}\xi_{a}=2\left(\frac{t}{ r}\right)\left(\eta_{ij}-\frac{x_{i}x_{j}}{r^{2}}\right)\delta^{i}_{a}\delta^{j}_{b}\;, \label{Keqforu}\eeq
where the $\delta^{i}_{a}\delta^{j}_{b}$ are present to project the non-zero contributions. We see that $\xi^{a}$ is a vector which satisfies Killing's equation in specific metric components, and one which fails as a modified CKV in other components. This comparison leads us to define a conformal factor associated with $\xi$:
\beq \Omega_{\xi}\equiv\frac{1}{(D-2)}\nabla_{c}\xi^{c}=\frac{t}{ r}\;,\eeq
for which one finds 
\beq \nabla_{d}\Omega_{\xi}=-\frac{1}{r^{2}}\xi_{d}\;,\quad N_{\xi}^{-1}\equiv||\nabla_{a}\Omega_{\xi}||=\frac{\alpha}{r^{2}}\;,\eeq
and
\beq u_{a}=N_{\xi}\nabla_{a}\Omega_{\xi}\;.\eeq
It is also straightforward to work out
\beq \nabla_{d}(\mathcal{L}_{\xi}g_{ab})|_{t=0}=\frac{2}{N_{\xi}}u_{d}\delta^{i}_{a}\delta^{j}_{b}\left(\eta_{ij}-\frac{x_{i}x_{j}}{r^{2}}\right)\;, \label{nabLg}\eeq
and 
\beq K_{\partial\Sigma}=\frac{1}{\alpha}(D-2)\;,\eeq
where $\mathcal{L}_{\xi}$ is the Lie derivative along $\xi_{a}$, and the extrinsic curvature of the spherical boundary $\partial\Sigma$ is $K=h^{ab}K_{ab}=g^{ab}\nabla_{b}n_{a}$, since $h_{ab}=g_{ab}-n_{a}n_{b}$.

\hspace{2mm}


\section{Thermodynamics of Causal Diamonds} 
\label{sec:thermoCD}
\indent

Consider the past of the causal diamond, i.e., the bottom half below the $t=0$ co-dimension-2 spherical slice $\partial B$ of Fig. 2\footnote{We focus on the past of the causal diamond for reasons which we will discuss in the discussion.}. Our picture for a physical process will be comparing the entropy between a time slice at $t=-\epsilon$ for positive $\epsilon$ and $t=0$ after some energy flux has entered the past of the diamond. At the boundary $t=\ell\pm r$, $\zeta^{2}=0$, and therefore, in Minkowski space, the boundary of the causal diamond represents a conformal Killing horizon of constant surface gravity $\kappa$, and therefore an isothermal surface with Hawking temperature $T=\kappa/2\pi$. An arbitrary spacetime will include curvature corrections, however, to leading order in a RNC expansion about a point $p$, $\zeta^{2}\approx0$, and $\kappa$ remains approximately constant. If we followed the worldline of $\zeta$ from time $t=-\epsilon$ to $t=0$, we would find that $\kappa$ would be different at each of these time slices. Motivated by the set-up of the stretched lightcone, we choose a timescale $\epsilon\ll\ell$ over which the surface gravity $\kappa$ is approximately constant. Therefore, in an arbitrary spacetime $\partial B$ of the causal diamond represents a local conformal Killing horizon, which may be interpreted as an isothermal surface with constant Hawking temperature $T=\kappa/2\pi$. 

We associate with this conformal Killing horizon a gravitational entropy \cite{Nielsen:2017hxt}, i.e., time-slices $\partial B$ of the causal diamond have an attributed entropy. The form of the entropy depends on the theory of gravity under consideration, e.g., for Einstein gravity, the correct form is the Bekenstein-Hawking entropy (\ref{BHent}). Here we consider a diffeomorphism invariant theory of gravity in $D$ spacetime dimensions defined by the action $I$:
\beq I=\frac{1}{16\pi G}\int d^{D}x\sqrt{-g}L\left(g^{ab},R_{abcd}\right)+I_{\text{matter}}\;.\eeq
Here the gravitational Lagrangian $L$ is written as a function of the metric and the curvature tensor $R_{abcd}$. This action encompasses a large class of theories of gravity which do not involve the derivatives of the Riemann tensor, e.g., $f(R)$ gravity, and Lovelock theories of gravity. The equations of motion for such theories are
\beq P_{a}^{\;cde}R_{bcde}-2\nabla^{c}\nabla^{d}P_{acdb}-\frac{1}{2}Lg_{ab}=8\pi G T_{ab}\;.\label{eqnsofmotion}\eeq
It is straightforward to verify that in the case of Einstein gravity, $L=R$, this reduces to Einstein's field equations. 

For a general theory of gravity of this type we must generalize the Bekenstein-Hawking entropy formula. We take this generalization to be the Wald entropy \cite{Wald93-1}:
\beq S_{\text{Wald}}=\frac{1}{8G\kappa}\int dS_{ab}J^{ab}\;,\eeq
where we have introduced the Noether potential associated with a diffeomorphism $x^{a}\to x^{a}+\zeta^{a}$, where we will take $\zeta^{a}$ to be a timelike (conformal) Killing vector, 
\beq J^{ab}=-2P^{abcd}\nabla_{c}\zeta_{d}+4\zeta_{d}\nabla_{c}P^{abcd}\;,\quad P^{abcd}\equiv\frac{\partial L}{\partial R_{abcd}}\;,\eeq
and have infinitesimal binormal element of $\partial B$:
\beq dS_{ab}\equiv\frac{1}{2}(N_{a}U_{b}-N_{b}U_{a})dA=\frac{1}{2}N_{ba}dA\;.\eeq
Wald's Noether charge construction of gravitational entropy was originally developed to yield an expression for the entropy of a stationary black hole in more general theories of gravity. Here we make the non-trivial assumption of local holography that this gravitational entropy can also be attributed locally to the spatial sections of causal diamonds whose structure is preserved by $\zeta_{a}$. 

For computational convenience, we will first not work directly on the horizon, but instead work on the timelike stretched horizon of the causal diamond -- a co-dimension-1 timelike surface we call $\Sigma$. At the end of the calculation we will take the limit where our stretched horizon coincides with the conformal Killing horizon. The fact that we have to take the step in which we move to the conformal  Killing horizon -- a null hypersurface -- is a marked difference with the analogous calculation using stretched future lightcones \cite{Parikh:2017aas}. 

The Wald entropy at time $t$ is
\beq S_{\text{Wald}}=-\frac{1}{4G\kappa}\int_{\partial B(t)}dS_{ab}(P^{abcd}\nabla_{c}\zeta_{d}-2\zeta_{d}\nabla_{c}P^{abcd})\;.\eeq
The total change in entropy between $t=0$ and $t=-\epsilon$ is $\Delta S_{\text{Wald}}=S_{\text{Wald}}(0)-S_{\text{Wald}}(-\epsilon)$, or, 
\beq \Delta S_{\text{Wald}}=\pm\frac{1}{4G\kappa}\int_{\Sigma}d\Sigma_{a}\nabla_{b}(P^{abcd}\nabla_{c}\zeta_{d}-2\zeta_{d}\nabla_{c}P^{abcd})\;, \label{DeltaS1}\eeq
where we have invoked Stokes' theorem for an antisymmetric tensor field $M^{ab}$:
\beq \int_{\Sigma}d\Sigma_{a}\nabla_{b}M^{ab}=\pm\left[\int_{\partial B(0)}dS_{ab}M^{ab}-\int_{\partial B(-\epsilon)}dS_{ab}M^{ab}\right]\;,\eeq
where the overall sign depends on whether $\Sigma$ is timelike ($-$), or spacelike ($+$). For our discussion of causal diamond thermodynamics we are interested in the timelike version, however, it will be illustrative for future discussion if we do not specify, for now, the signature of co-dimension-1 surface $\Sigma$. 

Moving on, we have
\beq 
\begin{split}
\Delta S_{\text{Wald}}&=\pm\frac{1}{4G\kappa}\int_{\Sigma}d\Sigma_{a}\{-\nabla_{b}(P^{adbc}+P^{acbd})\nabla_{c}\zeta_{d}\\
&+P^{abcd}\nabla_{b}\nabla_{c}\zeta_{d}-2\zeta_{d}\nabla_{b}\nabla_{c}P^{abcd}\}\;.
\end{split}
\eeq
We have yet to use any properties of $\zeta_{d}$, which to leading order is a conformal Killing vector, satisfying (\ref{confKeqn}) and  (\ref{confKid}). We have then:
\beq 
\begin{split}
\nabla_{b}(P^{adbc}+P^{acbd})\nabla_{c}\zeta_{d}&=\nabla_{b}P^{adbc}(\nabla_{c}\zeta_{d}+\nabla_{d}\zeta_{c})\\
&=2\Omega g_{cd}\nabla_{b}P^{adbc}\;,
\end{split}
\label{nabPnabz}\eeq
and
\beq
\begin{split}
 P^{abcd}\nabla_{b}\nabla_{c}\zeta_{d}&=P^{abcd}[R_{ebcd}\zeta^{e}+(\nabla_{c}\Omega)g_{bd}-(\nabla_{d}\Omega)g_{bc}]\\
&=P^{abcd}R_{ebcd}\zeta^{e}+2P^{abcd}(\nabla_{c}\Omega)g_{bd}\;,
\end{split}
\label{Pnabnabz}
\eeq
where we used that $P^{abcd}$ shares the same algebraic symmetries of the Riemann tensor. Substituting (\ref{nabPnabz}) and (\ref{Pnabnabz}) into (\ref{DeltaS1}) yields
\beq 
\begin{split}
\Delta S_{Wald}&=\pm\frac{1}{4G\kappa}\int_{\Sigma}d\Sigma_{a}\{P^{abcd}R_{ebcd}\zeta^{e}-2\zeta_{d}\nabla_{b}\nabla_{c}P^{abcd}\\
&+2P^{abcd}(\nabla_{c}\Omega)g_{bd}-2\Omega g_{cd}\nabla_{b}P^{adbc}\}\;,
\end{split}
\label{DeltaS2}\eeq
where the overall $+$ ($-$) sign indicates that $\Sigma$ is a timelike (spacelike) surface. In appendix (\ref{sec:appendA}), we consider the spacelike surface and provide an alternative derivation to the first law of causal diamond mechanics for higher derivative theories of gravity as presented in \cite{Bueno16-1}. 

Using that $d\Sigma_{a}=N_{a}dAd\tau=\partial_{a}^{r}dAd\tau=x_{i}/r\partial_{a}^{i}dAd\tau$, and that we are integrating over a spherically symmetric  region, we find that to leading order in the RNC expansion, that the final two terms integrate to zero since we are integrating over a timelike surface with spherical compact sections. Thus, to leading order, 
\beq \Delta S_{\text{Wald}}\approx\frac{1}{4G\kappa}\int_{\Sigma}d\Sigma_{a}(P^{abcd}R_{ebcd}\zeta^{e}-2\zeta_{d}\nabla_{b}\nabla_{c}P^{abcd})\;.\label{DeltaStotlead}\eeq
The two terms we neglect here, of course, have higher order contributions due to the RNC expansion, and in order to derive the non-linear equations of motion we must deal with these higher order contributions. We follow the technique developed in \cite{Parikh:2017aas}, in which we modify the conformal Killing vector $\zeta_{a}$ by adding $\mathcal{O}(x^{3})$ corrections and higher such that they remove the undesired higher order effects of the two terms we neglect. The details may be found in the appendix (\ref{sec:appendB}). 

The above expression (\ref{DeltaStotlead}) represents the leading order contribution to the total entropy variation, including the effect due to the natural increase of the spatial sections of the (past) causal diamond -- an irreversible thermodynamic process. Presently we are interested in the change in entropy due to a flux of matter crossing the conformal horizon -- a reversible thermodynamic process\footnote{We can consider the following analogy to help describe this process and our use of the terms `irreversible' and reversible': Imagine we have a box a gas sitting on a burner. When the box opens the gas will leave the box simply due to a free expansion, which has an associated irreversible entropy increase. The heating of the box will also lead to a reversible entropy increase. The natural increase of our diamond -- to the past of $t=0$ -- is analogous to the free expansion of the gas and we therefore identify this process as having an associated irreversible entropy increase.}. We therefore remove the entropy due to the natural increase of the diamond $\bar{S}$:
\beq
\begin{split} 
\bar{S}&=-\frac{1}{4G\kappa}\int_{\partial B}dAN_{i}U_{t}\left[P^{ittj}2\partial_{t}\zeta_{j}+P^{tijk}\partial_{j}\zeta_{k}\right]\\
&=\frac{1}{4G\kappa}\int_{\partial B}dA\frac{4}{r\ell^{2}}x_{i}x_{j}P^{ittj}\\
&=\frac{1}{4G\kappa}\frac{2\kappa K}{(D-2)}\frac{1}{(D-1)}\left(\sum_{i}P^{itti}\right)\Omega_{D-2}r^{D-1}\;,
\end{split}
\eeq
where to get to the second line we used that $\partial_{i}\zeta_{j}\propto\delta_{ij}$, which cancels with its contraction with $P^{tijk}$, and $\partial_{t}\zeta_{j}=-2x_{j}/\ell^{2}$, and in the third line we used that $2/\ell^{2}=\kappa K/(D-2)$, and again the fact we are integrating over a spherical subregion. To this order $P^{abcd}$ is constant, allowing us to pull it through the integral.

We may arrange the above suggestively as\footnote{As written, $\bar{S}$ is a bit misleading. It would appear that $\bar{S}$ goes like the volume rather than the area. However, this is in fact not the case. Indeed, in the case of general relativity, using $K=(D-2)/\ell$, and that on the $t=0$ slice $\partial B$, $r=\ell$, it is straightforward to show that $\bar{S}=A/4G$, where $A$ is the area of the spherical subregion $\partial B$.}\footnote{In the context of general relativity, we note that the this expression is nothing more than the Smarr formula for a maximally symmetric ball in flat space -- the ``thermodynamic volume" is notably absent \cite{Jacobson:2018ahi}. This is because we are considering perturbations about Minkowski spacetime. Even if we considered perturbations about a more general MSS, the thermodynamic volume would be subdominant.}
\beq \bar{S}=\frac{1}{2G}\frac{K}{(D-2)}\left(\sum_{i}P^{itti}\right)\int_{B}dV\;.\eeq
This expression is recognized to be the leading contribution  of the \emph{generalized volume} $\bar{W}$ (\ref{Wbar})
\beq
\begin{split}
\frac{K}{2G(D-2)}\int_{B}dVP^{abcd}U_{a}U_{d}h_{bc}\equiv\frac{K}{2G}\bar{W}\;,
\end{split}
\eeq
that is, 
\beq\Delta\bar{S}=\frac{K}{2G}\Delta\bar{W}\;,\eeq
where $\Delta \bar{S}=\bar{S}(0)-\bar{S}(-\epsilon)$, and $\Delta \bar{W}=\bar{W}(0)-\bar{W}(-\epsilon)$. Since the area on a future time slice $\partial B(0)$ is smaller than the that of $\partial B(-\epsilon)$, one has $\Delta\bar{S}>0$. Note that this is not the case for time-slices to the future of $t=0$, and therefore the thermodynamics of causal diamonds is peculiar; we will have more to say about this in the discussion. 

We therefore define the reversible entropy variation as
\beq
\begin{split}
 \Delta S_{\text{rev}}&\equiv\Delta S_{\text{Wald}}-(\Delta\bar{S})=\Delta S_{\text{Wald}}-\frac{K}{2G}\Delta\bar{W}\\
&=\frac{1}{4G\kappa}\int_{\Sigma}d\Sigma_{a}\left(P^{abcd}R_{ebcd}\zeta^{e}-2\zeta_{d}\nabla_{b}\nabla_{c}P^{abcd}\right)\;.
\end{split}
\label{DSrev}
\eeq
Calling this variation the reversible change in entropy is analogous to the Clausius relation in ordinary thermodynamics $Q=T\Delta S_{\text{rev}}$. 

\subsection{Gravity From Thermodynamics}
\indent

Next, following \cite{Jacobson95-1,Parikh:2017aas}, define the integrated energy flux across $\Sigma$ as 
\beq Q=\int_{\Sigma}d\Sigma_{a}T^{ab}\zeta_{b}\;,\eeq
where the energy momentum tensor can be approximated to leading order by its value at $p$. As we make the transition to the conformal Killing horizon, the interior of $\Sigma$ becomes causally disconnected from its exterior, allowing us to identify $Q$ as heat -- energy which flows into macroscopically unobservable degrees of freedom. 

The Clausius relation $T\Delta S_{\text{rev}}=Q$ for our set-up results in the geometric constraint:
\beq
\begin{split}
&\int_{\Sigma}d\Sigma_{a}\left(P^{abcd}R_{ebcd}\zeta^{e}-2\zeta_{d}\nabla_{b}\nabla_{c}P^{abcd}\right)\\
&=8\pi G\int_{\Sigma}d\Sigma_{a}T^{ab}\zeta_{b}\;.
\end{split}\eeq
Since this holds for all causal diamonds $\Sigma$, we may equate the integrands leading to
\beq (P^{aecd}R_{becd}-2\nabla^{d}\nabla^{c}P_{abcb})N^{a}\zeta^{b}=8\pi GT_{ab}N^{a}\zeta^{b}\;.\eeq
At the boundary, $t=\ell+r$, i.e., when the timelike stretched surface moves to the conformal Killing horizon, one has $g_{ab}N^{a}\zeta^{b}=0$. Therefore, at the conformal Killing horizon, the above is valid up to a term of the form $f g_{ab}$, where $f$ is some yet to be determined scalar function. The form of $f$ can be determined by demanding covariant conservation of $T_{ab}$. Specifically, we are led to 
\beq P^{aecd}R_{becd}-2\nabla^{d}\nabla^{c}P_{abcb}-\frac{1}{2}Lg_{ab}+\Lambda g_{ab}=8\pi GT_{ab}\;,\eeq
where $L(g^{ab},R_{abcd})$, and $\Lambda$ is some integration constant. We recognize the above as the equations of motion for a general theory of gravity. In this way we see that the equations of motion for a theory of gravity arise from the thermodynamics of causal diamonds. We have reproduced the results of \cite{Parikh:2017aas}, however, using the geometric construction of causal diamonds. 

This approach to deriving the equations of motion offers a thermodynamic perspective to the derivation of linearized equations of motion from the entanglement equilibrium proposal as presented in \cite{Bueno16-1}. In particular, we found that the generalized volume $\bar{W}$ can be interpreted as the natural increase of the causal diamond. To apply the Clausius relation for a reversible thermodynamic process, we removed this increase and, therefore, $\bar{W}$ is the contribution which generates irreversible thermodynamic processes in the causal diamond construction. We note that removing $\bar{W}$ also appears in the first law of causal diamond mechanics (\ref{1stlawcdoff}), and consequently the entanglement equilibrium condition (\ref{1stlawEEmod}).

It is interesting to compare the above construction with that of the stretched future lightcone. As shown in \cite{Parikh:2017aas}, the non-linear equations of motion for the same class of theories of gravity arise as a consequence of the Clausius relation applied to the stretched future lightcone -- a co-dimension-1 timelike hyperboloid. Unlike the above derivation, one need not take the limit that the stretched horizon goes to a null surface. This is because the stretched horizon of the future lightcone acts as a causal barrier between observers living on the exterior of the cone from its interior, allowing for a well-defined notion of heat even in the absence of a Killing horizon. In the causal diamond set-up we had to take the limit that the stretched horizon moves to the conformal Killing horizon for technical reasons; it is unclear what the physical reason for this may be as the energy passing through the past causal diamond seemingly has a well-defined notion of heat. 

Moreover, in the future stretched light cone set-up, one similarly removes the entropy change due to the natural expansion of the hyperboloid. In light of the result above, that the entropy change due to the natural increase in the diamond may be interpreted as the generalized volume, naively we guess that the natural entropy change of the hyperoloid might have a similar interpretation. This suggests that we can think about the derivation of the gravitational equations of motion using the stretched future lightcone construction from an entanglement entropy perspective, i.e., perhaps the gravitational equatons of motion arise from an entanglement equilibrium condition, analogous to that given in \cite{Jacobson16-1,Bueno16-1}. We explore this idea in the next section.


\section{Entanglement of Lightcones} 
\label{sec:entLC}
\indent

Recently the equations of motion for a generalized theory of gravity were derived from the thermodynamics of the stretched future lightcone \cite{Parikh:2017aas}. Of course, thermodynamics is a placeholder until a more precise quantum prescription of a system is developed -- in this case, a quantum theory of spacetime. We can make progress, however, following the recent paradigm relating entanglement to geometry. Indeed, it is natural to interpret the thermodynamic entropy of the stretched lightcone as coming from the entanglement of quantum fields outside of the stretched horizon. Moreover, due to the geometric similarity between causal diamonds and stretched lightcones (\ref{sec:geosetup}), we are motivated by the derivation of linearized gravitational equations from the entanglement properties of the causal diamond \cite{Jacobson16-1,Bueno16-1}. We therefore aim to derive linearized equations of motion via the entanglement of the stretched lightcone. 

Our procedure is as follows. First we compute $\delta S_{\text{Wald}}$ and derive an off-shell geometric identity analogous to the first law of causal diamonds, which we call the first law of stretched lightcones. We will use the Noetheresque approach illustrated in (\ref{sec:appendA}). Next we will show how this off-shell identity is equivalent to the variation of the entanglement entropy, following arguments presented in \cite{Bueno16-1}. Finally, we will find that the linearized form of the gravitational equations emerge from an entanglement equilibrium condition. In essence, we are simply considering Jacobson's entanglement equilibrium proposal \cite{Jacobson16-1} for the geometry of stretched lightcones in an arbitrary background (where we explicitly consider perturbations to Minkowski space). One expects to find a similar result as established in \cite{Bueno16-1}, simply by noting that the stretched lightcone shares enough geometric similarities to the causal diamond. 

\subsection{First Law of Stretched Lightcones}
\indent

Recall that $\xi_{a}$ satisfies (\ref{Keqforu})
\beq \nabla_{a}\xi_{b}+\nabla_{b}\xi_{a}=2\Omega_{\xi}\tilde{g}_{ab}\;,\eeq
where $\Omega_{\xi}=t/ r$, and we have defined
\beq \tilde{g}_{ab}=\left(\delta_{ij}-\frac{x_{i}x_{j}}{r^{2}}\right)\delta^{i}_{a}\delta^{j}_{b}\;.\eeq
The derivation presented in (\ref{sec:appendA}) relies on the fact that $\zeta_{a}$ is an exact conformal Killing vector in flat space; specifically the fact that $\zeta_{a}$ satisfies the conformal Killing identity.  Here the vector $\xi_{a}$ is not a conformal Killing vector, and therefore, it will not satisfy the conformal Killing identity. The issue is that $\tilde{g}_{ab}$ defined above is not the metric, and therefore this object will have a non-vanishing covariant derivative. However, since we are considering the time $t=0$ surface, the fact that $\xi_{a}$ does not satisfy the conformal Killing identity is not a problem for us because $\Omega_{\xi}$ will vanish at $t=0$. Therefore, all terms $\Omega_{\xi}\nabla\tilde{g}$ which would appear can be neglected. 

Therefore, we may simply start from (\ref{DeltaS2}) above with the following substitutions
\beq \zeta^{a}\to \xi^{a}\;,\quad\Omega\to\Omega_{\xi}\;,\quad g_{ab}\to\tilde{g}_{ab}\;.\eeq
Moreover, for the lightcone $\kappa=1$, and on the co-dimension-1 spatial ball $B$, $\alpha$ is set to be constant\footnote{The acceleration of the spherical Rindler observers is fixed as a constant at $t=0$, via the construction described in (\ref{sec:geosetup}).}, and has volume element $dB_{a}=U_{a}dV$. Thus, we have that
\beq 
\begin{split}
 S_{Wald}&=-\frac{1}{4G}\int_{B}dB_{a}\{P^{abcd}R_{ebcd}\xi^{e}-2\xi_{d}\nabla_{b}\nabla_{c}P^{abcd}\\
&+2P^{abcd}(\nabla_{c}\Omega_{\xi})\tilde{g}_{bd}\}\;,
\end{split}
\label{deltaSLC}\eeq
where we dropped the term proportional to $\Omega_{\xi}$ as it vanishes at $t=0$. 

Let us study the bottom line. Using $(\nabla_{c}\Omega_{\xi})|_{t=0}=-1/r^{2}\xi_{c}$, we find to leading order we have
\beq
\begin{split}
&-\frac{1}{4G}\int_{B}dB_{a}2P^{abcd}(\nabla_{c}\Omega_{\xi})\tilde{g}_{bd}\\
&=-\frac{1}{2G}\int_{B}dV\frac{P^{tijt}}{r}\left(\delta_{ij}-\frac{x_{i}x_{j}}{r^{2}}\right)\\
&=-\frac{1}{2G}\frac{1}{(D-1)}\left(\sum_{i}P^{tiit}\right)\Omega_{D-2}r^{D-2}\;.
\end{split}
\eeq
Note that this object is proportional to the surface area of the spherical subregions; in fact in the case of Einstein gravity, $P_{GR}^{abcd}=\frac{1}{2}(g^{ac}g^{bd}-g^{ad}g^{bc})$, the above simply becomes $-\frac{A_{\partial B}}{4G}$, the Bekenstein-Hawking entropy. Motivated by the derivation of the first law of causal diamonds in \cite{Bueno16-1} we might be inclined to refer to this object as the \emph{generalized area}\footnote{In fact, we could also interpret this quantity as being proportional to the generalized volume. Using $K_{\partial\Sigma}=(D-2)/\alpha$, and that we are integrating a ball of radius $\alpha$, we find that this term may be expressed as $K/2G\bar{W}$.}, however, this object appears in \cite{Parikh:2017aas} (see equations (67)-(68) of their paper), and is identified as the entropy due to the natural background expansion of the hyperboloid, $\bar{S}$. Specifically, 
\beq  \bar{S}=-\frac{1}{4G}\int_{B}dB_{a}2P^{abcd}(\nabla_{c}\Omega_{\xi})\tilde{g}_{bd}\;,\label{genarea}\eeq
and therefore, 
\beq 
\begin{split}
 S_{\text{Wald}}- \bar{S}&=-\frac{1}{4G}\int_{B}dB_{a}\{P^{abcd}R_{ebcd}\xi^{e}\\
&-2\xi_{d}\nabla_{b}\nabla_{c}P^{abcd}\}\;.
\end{split}\eeq

Next, introduce the matter energy $H^{m}_{u}$ associated with spherical Rindler observers with proper velocity $u$, 
\beq H^{m}_{u}=\int_{B}dB_{a}T^{ab}u_{b}\;.\eeq
Then, following the same arguments given in \cite{Jacobson16-1,Bueno16-1} (and reviewed in (\ref{sec:appendA})), we find
\beq \frac{1}{2\pi\alpha}(\delta S_{\text{Wald}}-\delta \bar{S})=-\delta H_{u}^{m}\; \label{firstlawstretchedLCon}\eeq
is equivalent to the linearized gravitational  equations of motion about flat spacetime for $L(g^{ab},R_{abcd})$ theories of gravity:
\beq \delta G^{ad}-2\partial_{b}\partial_{c}(\delta P^{abcd}_{\text{higher}})=8\pi G\delta T^{ad}\;.\eeq

The off-shell identity is simply 
\beq \frac{1}{2\pi\alpha}(\delta S_{\text{Wald}}-\delta \bar{S})+\delta H^{m}_{u}=\int_{B}\delta C_{\xi}\;,\label{firstlawstretchedLCoff}\eeq
where $\delta C_{\xi}$ represents the linearized constraint that the gravitational field equations hold. 

We can actually understand this first law of stretched lightcones as the Iyer-Wald identity \cite{Iyer:1994ys} in the case of the stretched horizon of spherical Rindler observers, rather than the dynamical horizon of a black hole. As illustrated in (\ref{sec:appendC}), we may actually interpret the generalized area as the variation of the gravitational Hamiltonian. 

Moreover, the first two terms on the LHS of (\ref{firstlawstretchedLCoff}) can be combined into a single object  \cite{Bueno16-1}, namely, the variation of the Wald entropy while keeping the generalized area constant, i.e., 
\beq \frac{1}{2\pi\alpha}(\delta S_{\text{Wald}}-\delta \bar{S})=\frac{1}{2\pi\alpha}\delta S_{\text{Wald}}|_{\bar{S}}\;,\eeq
leading to
\beq \frac{1}{2\pi\alpha}\delta S_{\text{Wald}}|_{\bar{S}}+\delta H^{m}_{u}=\int_{B}\delta C_{\xi}\;.\eeq

The Wald formalism contains the so-called JKM ambiguities \cite{Jacobson:1993vj}; one may add an exact form $dY$ linear in the field variations and their derivatives to the Noether current, and $Y$ to the Noether charge. This would lead to a modification of $S_{\text{Wald}}$ and $\bar{S}$. However, it is clear the combined modification will cancel, allowing us to write
\beq \frac{1}{2\pi\alpha}\delta S_{\text{Wald}}|_{\bar{S}}=\frac{1}{2\pi\alpha}\delta(S_{\text{Wald}}+S_{JKM})|_{\bar{S}'}\;,\eeq
where $\bar{S}'=\bar{S}+\bar{S}_{JKM}$. For more details on this calculation one need only follow the calculation presented in \cite{Bueno16-1} as it is identical in the stretched lightcone geometry. 

\subsection{Gravity from Entanglement}
\indent

Our aim here is to show how the first law of stretched lightcones -- an off-shell geometric identity -- can be understood as a condition on entanglement entropy. Before we consider the scenario with stretched lightcones, let us recall what happens in the case of a causal diamond. The entanglement equilibrium conjecture makes four central assumptions which we outline here and are reviewed in (\ref{sec:appendA}). These assumptions include \cite{Carroll:2016lku}: (i) Entanglement separability, i.e., $S_{EE}=S_{UV}+S_{IR}$; (ii) equilibrium condition, i.e., a simultaneous variation of the quantum state and geometry of the entanglement entropy of the causal diamond is extremal, and the geometry of the causal diamond is that of a MSS; (iii) Wald entropy as UV entropy, i.e., the variation of the UV entropy is proportional to the Wald entropy at fixed generalized volume, and (iv) CFT form of modular energy, i.e., the modular energy is defined to be the variation of the expectation value of the modular Hamiltonian -- which for spherical regions may be identified with the Hamiltonian generating the flow along the CKV which preserves the causal diamond -- plus some scalar operator $X$. 

Reference \cite{Jacobson16-1} showed that the above postulates can be used to derive the full non-linear Einstein equations, while \cite{Bueno16-1} showed these postulates lead to the linearized gravitational equations for higher derivative theories of gravity. Here we will discuss how to justify the above assumptions (for a more pedagogical review, see \cite{Carroll:2016lku}) and attempt to apply a similar set of assumptions for the case of stretched lightcones. 

Assumption (i), where we require minimal entanglement between IR and UV degrees of freedom, is in fact a fundamental feature of renormalization group (RG) flows. More precisely, an RG flow requires a decoupling between high and low momentum states. Thus, in a Wilsonian effective action we would expect minimal entanglement between UV and IR modes. We also would assume that this basic feature of effective field theory to continue to hold in the theory's UV completion. This assumption is reasonably justified in both the causal diamond and stretched lightcone set-ups. 

The second assumption (ii) asserts that the vacuum state in a small region of spacetime may be described by a Gibb's energy state, and that for a fixed energy, this state will have a maximum entropy, i.e., $\delta S_{EE}=0$. Moreover, the requirement that the causal diamond is described in a MSS is simply there to prevent curvature fluctuations from producing a large backreaction which spoil the equilibrium condition. In other words, the semiclassical (linearized) equations hold if and only if the causal diamond is in thermodynamic equilibrium. Likewise, we may safely make this same assumption about the stretched lightcone: when the stretched lightcone is in thermal equilibrium, the gravitational equations hold (via the Clausius relation), and vice versa. 

Assumption (iii), like assumption (i), is also not very controversial. All that is being said is that one should identify the area $\partial B$ of the causal diamond, and, similarly, the cross-sectional area of the stretched lightcone $\partial \Sigma$, as the area of the planar Rindler horizons existing at the edge of the causal diamond, and the area of the timelike spherical Rindler horizon, respectively. Motivated by the Ryu-Takayanagi proposal, we then simply identify these areas with the entanglement entropy of each region. We should point out a difference between the two pictures, however. It is known that the entanglement entropy of the causal diamond $D[B]$, i.e., the causal domain of a spherical ball region $B$, is equivalent to the entanglement entropy of $B$ itself. Meanwhile, we are saying that the entanglement entropy of the stretched horizon, $\Sigma$, is equivalent to the ball $B$ whose boundary is $\partial\Sigma$. This has been established in the context of spherical Rindler space, which we may interpret our stretched lightcone as being: The entanglement entropy of spherical Rindler space is equal to the area of the horizon $\partial\Sigma$ \cite{Balasubramanian13-1}. 

Unlike the first three assumptions, which all rely on the underlying UV physics, assumption (iv) makes an assertion about the form of the modular Hamiltonian for IR degrees of freedom. In the case of causal diamonds one makes two observations. First, a causal diamond in Minkowski space may be conformally transformed to a (planar) Rindler wedge. Then, via an application of the Bisognano-Wichmann theorem \cite{Bisognano:1976za}, for CFTs the modular Hamiltonian $H_{\text{mod}}$, defined via the thermal state $\rho_{IR}=Z^{-1}e^{-H_{\text{mod}}}$, is proportional to the Hamiltonian generating the flow along the CKV $\zeta$, i.e., $H_{\text{mod}}=2\pi/\kappa H^{m}_{\zeta}$ \cite{Casini:2011kv}. This implies then that the variation of the modular Hamiltonian is equal to the variation of of $H^{m}_{\zeta}$, plus some additional spacetime scalar $X$, i.e., 
\beq \delta\langle H_{\text{mod}}\rangle=\frac{2\pi}{\kappa}\delta\int_{B}dB_{a}(T^{ab}\zeta_{b}+Xg^{ab}\zeta_{b})\;.\eeq
This specific assumption is interesting in that it may be explicitly checked, and has been justified \cite{Casini:2016rwj,Carroll:2016lku}, though with the stipulation that $X$ may depend on $\ell$. 

In the case of stretched lightcones, our assumption is then that the modular Hamiltonian $H^{\text{mod}}_{u}$, defined by $\rho_{\Sigma}=Z^{-1}e^{-H_{\text{mod}}}$, is proportional to the radial boost Hamiltonian,
\beq H_{\text{mod}}=2\pi\alpha\int_{B}dB_{a}T^{ab}u_{b}\;,\eeq
and that we may also include a spacetime scalar $X$. We would like to be able to similarly justify this assumption, as was accomplished in the causal diamond case. While currently this assumption is non-trivial and has not been computationally justified, we find that it is reasonable, as we now describe. 

The stretched lightcone $\Sigma$, like spherical Rindler space, can be understood as the union of Rindler planes; indeed, if we constrain ourselves to the $y=z=0$ plane, the radial boost vector $\xi^{a}=r\delta^{a}_{t}+t\partial^{a}_{r}$ reduces to a Cartesian boost vector. Each Rindler plane may be associated with a single causal diamond. The union of these causal diamonds yields a single ``radial causal diamond" \cite{Balasubramanian13-1}\footnote{This is precisely the construction of spherical Rindler space. If we were to embed spherical Rindler space into AdS, i.e., spherical Rindler-AdS space, the radial causal diamond was found to be holographically dual to a finite time strip in a boundary field theory \cite{Balasubramanian:2013lsa}.}. Therefore, the congruence of uniformly and constantly, radially accelerating observers comprising the stretched lightcone have an associated radial causal diamond. Moreover, the radial boost $\xi^{a}$ preserves the flow of the hyperboloid $\Sigma$. Our assumption is that the entanglement entropy of the stretched lightcone is that of the radial causal diamond which is also that of spherical region $B$. Thus we define the modular Hamiltonian as above and assume that it is proportional to the Hamiltonian generating the flow of $\Sigma$. For similar arguments given in \cite{Casini:2016rwj,Carroll:2016lku}, we expect -- but have not proved -- that for CFTs we may also modify the modular Hamiltonian by a spacetime scalar.

Let us now briefly show how the first law of stretched lightcones -- an off-shell geometric identity -- can be understood as a condition on entanglement entropy. In particular, we can follow the discussion given in \cite{Bueno16-1} (also reviewed in (\ref{sec:appendA}), making only a few simple changes). We perform a simultaneous (infinitesimal) variation of the entanglement entropy on a stretched lightcone of $S_{EE}$ with respect to the geometry and quantum state. By entanglement separability, $\delta S_{EE}$ takes the form
\beq \delta S_{EE}=\delta S_{UV}+\delta S_{IR}\;,\eeq
where the UV contribution is state independent and is assumed to be given by $\delta S_{UV}=\delta (S_{\text{Wald}}+S_{JKM})_{\bar{S}'}$, while the IR contribution comes from the modular Hamiltonian via the first law of EE, $\delta S_{IR}=\delta\langle H_{\text{mod}}\rangle=2\pi\alpha\delta\langle H^{m}_{u}\rangle$. 
Then, using the first law of entanglement entropy for a system in which the background geometry is also varied (\ref{1stlawEEmod}), we arrive to
\beq \frac{1}{2\pi\alpha}\delta S_{EE}|_{\bar{S}'}=\int_{B}\delta C_{\xi}\;,\eeq
valid for minimally coupled, conformally invariant matter fields. 

Thus, there is an equivalence between the following statements: (i) $S_{EE}$ is maximal in vacuum for all balls in all frames, and (ii) the linearized higher derivative equations hold everywhere. In other words, the entanglement equilibrium condition is equivalent to the linearized higher derivative equations of motion to be satisfied, and vice versa. This equivalence may be verified via a simple modification of the calculations presented in \cite{Bueno16-1}. We also note that here we considered perturbations about Minkowski space, however, one could, in principle, generalize this to a MSS, and while the above discussion was particular to theories of gravity described by $L(g^{ab},R_{abcd})$, i.e., those which do not depend on the derivatives of the Riemann tensor, we could have included those derivatives as well.


\section{Discussion and Future Work}
\indent

After reviewing the geometric similarities between causal diamonds and stretched future lightcones (\ref{sec:geosetup}), we presented a derivation of the full non-linear gravitational equations of motion in (\ref{sec:thermoCD}) by assigning thermodynamics to the conformal Killing horizon of the causal diamond, i.e., a Hawking temperature $T_{H}=\kappa/2\pi$, and a local holographic entropy $S_{\text{Wald}}$. The equations of motion were a geometric consequence of the Clausius relation $T_{H}\Delta S_{\text{rev}}=Q$, where $\Delta S_{\text{rev}}$  (\ref{DSrev}) is defined as the entropy solely due to a matter flux crossing the horizon. We found that the quantity $\frac{K}{2G}\bar{W}$, where $\bar{W}$ is the generalized volume, can be understood as the entropy of the natural increase of the causal diamond. This provides a microscopic interpretation of the generalized volume. Our physical process derivation of the equations of motion was motivated by \cite{Parikh:2017aas} where the gravitational equations were derived from the thermodynamics of stretched future lightcones. 

Motivated by \cite{Jacobson16-1,Bueno16-1}, in (\ref{sec:entLC}) we showed how to derive the linearized gravitational equations of motion from the entanglement equilibrium proposal, i.e., that the entanglement entropy for spherical entangling regions is maximal in the vacuum. We did this by first deriving an off-shell geometric identity, the first law of stretched lightcones, and showed that it was equivalent to the first law of entanglement entropy in the case of spherical subregions and conformally invariant matter. In the derivation of the first law of stretched lightcones we found an expression for the generalized area, which is nothing more than the entropy due to the natural expansion of the stretched lightcone. To complete this derivation, however, we to had make the non-trivial assumption that the entanglement entropy of the spherical entangling region $\partial\Sigma$ is the entanglement entropy of $\Sigma$, and the  modular Hamiltonian $H_{\text{mod}}$ is proportional to the radial boost Hamiltonian $H^{m}_{u}$.  This is a speculation which requires justification and will be persued in the future.


We can summarize our findings of (\ref{sec:entLC}) and the equivalent statement for causal diamonds \cite{Jacobson16-1,Bueno16-1} as
\beq T\delta S_{EE}|_{\bar{S}'}=\int_{B}\delta C\;,\eeq
where $\bar{S}'$ is the irreversible entropy due to the natural change of the background geometry -- identified as the generalized volume in the case of causal diamonds, or the generalized area in the case of stretched lightcones -- and where $T$ is the temperature associated with the horizon of the surface, namely, the Hawking temperature  $T_{H}=\kappa/2\pi$ in the case of causal diamonds, or the Unruh-Davies temperature $T=1/2\pi \alpha$ in the case of stretched lightcones. Entropy being maximal in the vacuum implies that the linearized constraint is satisfied, leading to the linearized form of the equations of motion of higher derivative theories of gravity, or, in the special case of Einstein gravity, the full non-linear equations. 

Apart from the entanglement equilibrium interpretation of the (mechanical) first laws of causal diamonds and stretched lightcones, it is natural to interpret them as thermodynamic relationships, namely we may view $T\delta S_{EE}|_{\bar{S}'}=0$ as the infinitesimal variation version of $T\Delta S_{\text{rev}}=Q$, where the condition of constant generalized volume (or generalized area) is equivalent to studying reversible thermodynamic processes. More specifically, the UV contribution to $\delta S_{EE}$ would be identified with change in reversible gravitational entropy $\Delta S_{\text{rev}}$, while the IR contribution to entanglement entropy would be identified with the heat $Q$.  Moreover, if entanglement entropy is UV finite, as assumed here, and satisfies the Clausius relation, then whatever the content of the underlying theory is, the entanglement entropy must be proportional to the Bekenstein-Hawking entropy, in the case the theory of spacetime is governed by Einstein gravity \cite{Jacobson:2012yt}.

Unlike the derivation of the non-linear equations using stretched lightcones \cite{Parikh:2017aas}, in the case of working with causal diamonds we had to take the limit that we are working on the conformal Killing horizon. One reason for this may be because the future stretched lightcone is defined as a surface of proper acceleration, and the binormal $dS_{ab}$ is constructed from $u_{a}$ -- the proper velocity -- and $n_{a}$ -- a normal vector proportional to the proper acceleration. In this way the stretched lightcone is directly analogous to the stretched horizon of a black hole, which has well-defined thermodynamics. Due to the geometric similarities, it appears that the (past) stretched causal diamond has a similar interpretation.

Another way in which the two physical process derivations are different is that in the case of stretched lightcones, $\Delta S_{\text{rev}}>0$ and $\Delta\bar{S}>0$. Therefore,  positive (classical) energy flux causes the (reversible) thermodynamic entropy to increase. Consequently, the null energy condition (NEC) is satisfied, and the stretched lightcone behaves as an ordinary thermodynamic system. These same features hold when we restrict ourselves to the past (before $t=0$) of the causal diamond. In contrast, the future of the causal diamond, i.e., to the future of $t=0$, one  has $\Delta\bar{S}<0$. From a geometric point of view, the reason for this decrease in thermodynamic entropy is clear: The cross-sectional area of the causal diamond is decreasing as one moves forward in time, reaching zero at the tip of the cone. Moreover, matter that is inside of the future of the diamond is free to leave the system -- there is no horizon preventing it from leaving, and matter entering from the outside must be moving faster than the speed of light\footnote{This is why we considered a heat flux entering the past of the causal diamond.}; there is even a question as to whether a diamond is thermodynamically stable\footnote{I would like to thank Ted Jacobson for pointing this observation out to me.}. In fact, the causal diamond has further non-classical thermodynamic properties: A causal diamond behaves as a system with negative temperature. However, in the context of entanglement equilibrium, a (conformal) matter flux yields a positive change in entanglement entropy $\delta S_{EE}>0$.  In a related context these observations were recently discussed in \cite{Jacobson:2018ahi}, where it is suggested that the ``classical" part of entropy is governed by negative temperature, while the quantum corrections present in the entanglement entropy are seemingly ruled by a positive temperature.

\subsection{Future Directions}
\indent

We conclude by noting ways in which our work may be extended. 

\subsubsection{Local First Laws}
\indent

We now have two derivations of the gravitational equations of motion via a thermodynamic process, and an application of the Clausius relation $T\Delta S_{\text{rev}}=Q$. Recently \cite{Parikh:2018anm}, it was shown that one may write down a hybrid first law of gravity and thermodynamics 
\beq \Delta E=T\Delta S_{\text{rev}}-\mathcal{W}\;,\label{firstlawlocgrav}\eeq
connecting matter energy $E$ and work $\mathcal{W}$ with the gravitational entropy $S$ evaluated on the stretched future lightcone of any point in an arbitrary spacetime. It would be interesting to see if we can find a similar first law of causal diamonds. In fact, recently, Jacobson has established a first law for a causal diamond in a maximally symmetric space, analogous to the first law of black hole mechanics \cite{Jacobson:2018ahi}. In this set-up, the causal diamond is equipped with a cosmological constant, and one discovers that a local gravitational first law of causal diamonds is reminiscent of the Smarr formula for a ball in a maximally symmetric space. Moreover, if one wishes to interpret this first law as a Clausius relation, then the causal diamond, classically, is a thermodynamic system with a negative temperature. It would be interesting to study the thermodynamic behavior of the causal diamond, as well as look for a similar local first law for stretched lightcones, and verify that the stretched lightcone is a thermodynamic system with positive temperature.

\subsubsection{Non-Linear Equations of Motion}
\indent

It is interesting that we were able to derive the full non-linear gravitational equations of motion via a reversible process, while we only found the linearized equations of motion via the entanglement equilibrium condition. This is because we restricted ourselves to first order perturbations of the entanglement entropy and background geometry.  Higher order perturbations to the entanglement entropy lead to a modified form of the first law of entanglement entropy, e.g., the second order change in entanglement entropy is no longer proportional to the expectation value of the modular Hamiltonian (\ref{firstlawEE}), but rather one must include the relative entropy. Moreover, as pointed out in \cite{Bueno16-1}, using higher order terms in the RNC expansion and higher order perturbations to the entanglement entropy could make it possible to derive the fully nonlinear equations of an arbitrary theory of gravity. Indeed, these ideas were recently incorporated in the context of holographic entanglement entropy to derive the non-linear contributions to gravitational equations \cite{Faulkner:2017tkh,Haehl:2017sot,Lewkowycz:2018sgn}. Due to the simlarity between the holographic and entanglement equlibrium approaches, developments in one is likely to inform the other. 

We should also point out that the way we derived the non-linear gravitational equations via a physical process was by modifying $\zeta_{a}$ and $\xi_{a}$ to deal with the fact that $\zeta_{a}$ and $\xi_{a}$ are both approximate Killing vectors. It would be interesting to see whether these modifications have a microscopic interpretation and could be employed in the context of entanglement equilibrium  such that the non-linear equations of motion arise without needing to consider second order perturbations to the entanglement entropy.

\bigskip
\noindent
{\bf Acknowledgments} \\
\noindent
I would like to thank Victoria Martin, Batoul Banihashemi, and Maulik Parikh for helpful comments on the manuscript, and Ted Jacobson for insightful discussions.

\appendix

\section{FLCD and Entanglement Equilibrium}
\label{sec:appendA}
\indent


\subsection{First Law of Causal Diamond Mechanics}
\indent

Here we present a slightly different derivation of the first law of causal diamond mechanics (FLCD) for higher derivative theories of gravity than given in \cite{Bueno16-1}. Let us take the minus sign of (\ref{DeltaS2}), when $\Sigma$ is the co-dimension-1 spacelike ball $B$. In this picture, the $\Delta$ is not referring to a comparison of $S_{\text{Wald}}$ at two different time slices, i.e., not a physical process -- all we have done is make use of Stokes' theorem. To make this point clear we drop the $\Delta$. 

 Following the same steps shown in (\ref{sec:thermoCD}), we have
\beq
\begin{split}  S_{\text{Wald}}&=-\frac{1}{4G\kappa}\int_{B}dB_{a}\{P^{abcd}R_{ebcd}\zeta^{e}-2\zeta_{d}\nabla_{b}\nabla_{c}P^{abcd}\\
&+2P^{abcd}(\nabla_{c}\Omega)g_{bd}-2\Omega g_{cd}\nabla_{b}P^{adbc}\}\;,
\end{split}
\eeq
where we have chosen to write the volume element of $B$ as $dB_{a}=U_{a}dV$. On $B(t=0)$, $\Omega=0$, leading to:
\beq 
\begin{split}
 S_{\text{Wald}}&=-\frac{1}{4G\kappa}\int_{B}dB_{a}\{P^{abcd}R_{ebcd}\zeta^{e}-2\zeta_{d}\nabla_{b}\nabla_{c}P^{abcd}\\
&+2P^{abcd}(\nabla_{c}\Omega)g_{bd}\}\;.
\end{split}
\label{deltaS1}\eeq
The final term is
\beq
\begin{split}
\frac{2K}{4G(D-2)}\int_{B}dVP^{abcd}U_{a}U_{d}h_{bc}\equiv\frac{K}{2G}\bar{W}\;,
\end{split}
\label{Wbar}\eeq
where we used $(\nabla_{c}\Omega)|_{B}=\kappa KU_{c}/(D-2)$, and introduced the induced metric $h_{bc}$ on $B$. This contribution $\bar{W}$ is proportional to a part of the \emph{generalized volume} introduced in \cite{Bueno16-1}: 
\beq W=\frac{1}{(D-2)P_{0}}\int_{B}dV(P^{abcd}U_{a}U_{d}h_{bc}-P_{0})\;.\label{Wvol}\eeq
 Here $P_{0}$ is a theory dependent constant defined by the $P^{abcd}$ tensor in a maximally symmetric solution to the field equations via $P^{abcd}_{MSS}=P_{0}(g^{ac}g^{bd}-g^{ad}g^{bc})$. It can be verified that in the case of Einstein gravity (\ref{Wvol}) is the spatial volume $V$ of the diamond. Our expression $\bar{W}$ does not include the $P_{0}$ term\footnote{We can arrive to the generalized volume (\ref{Wvol}) by subtracting $P^{abcd}_{MSS}$ from $P^{abcd}$ in the expression for the Wald entropy; specifically, replace $P^{abcd}$ with $P^{abcd}-\frac{1}{(D-1)}P^{abcd}_{MSS}$ in $S_{\text{Wald}}$. Repeating the steps that lead to (\ref{deltaS1}) will include an additional term which is precisely the extra term found in $W$, missing from $\bar{W}$.}.

 We observe that, like $W$, $\bar{W}$ is also proportional to the physical volume in the case of Einstein gravity. Specifically, in Einstein gravity, $P^{abcd}=1/2(g^{ac}g^{bd}-g^{ad}g^{bc})$, we find
\beq \bar{W}_{GR}=\frac{(D-1)}{(D-2)}V\;.\eeq
This expression is reminiscent of the Smarr formula for a maximally symmetric ball with a vanishing cosmological constant: $(D-2) A=(D-1)KV$ \cite{Jacobson:2018ahi}. This suggests that $\bar{W}$ is really related to the entropy; indeed, in the body of this report we will find such an interpretation when we study the thermodynamics of causal diamonds. 

Moving on, to linear order in the Riemann normal coordinate expansion, a perturbation about flat space leads to \cite{Bueno16-1}
\beq
\begin{split}
 &\delta \left(S_{Wald}-\frac{K}{2G}\bar{W}\right)\\
&=-\frac{U_{a}U_{d}}{4G\kappa}\int_{B}dV\left(P^{abcd}_{GR}\delta R^{d}_{\;bce}-2\partial_{b}\partial_{c}\delta P_{\text{higher}}^{abcd}\right)\left(1-\frac{r^{2}}{\ell^{2}}\right)\;,
\end{split}
\eeq
where we have separated $P^{abcd}=P^{abcd}_{GR}+P^{abcd}_{\text{higher}}$. Introducing the matter conformal Killing energy $H^{m}_{\zeta}$,
\beq H^{m}_{\zeta}=\int_{B}dVT_{ab}U^{a}\zeta^{b}\;, \label{matHam}\eeq
we find
\beq \delta H^{m}_{\zeta}=\int_{B}dV\delta T_{ab}U^{a}U^{b}\left(1-\frac{r^{2}}{\ell^{2}}\right)\;.\eeq
Notice then that for all timelike unit vectors one finds that 
\beq \frac{\kappa}{2\pi}\delta\left(S_{\text{Wald}}-\frac{K}{2G}\bar{W}\right)=-\delta H^{m}_{\zeta}\;,\label{FLCDnoether}\eeq
is equivalent to the tensor equation \cite{Jacobson16-1}:
\beq  \delta R^{ad}-2\partial_{b}\partial_{c}(\delta P^{abcd}_{\text{higher}})+(\delta X)\eta^{ad}=8\pi G\delta T^{ad}\;,\eeq
where we have introduced the spacetime scalar $X$, an assumption to be explained momentarily. Demanding local conservation of energy leads to 
\beq \delta\left(R^{ad}-\frac{1}{2}\eta^{ad}R+\Lambda \eta^{ad}\right)-2\partial_{b}\partial_{c}(\delta P^{abcd}_{\text{higher}})=8\pi G\delta T^{ad}\;,\label{lineareqnsmot}\eeq
which we recognize as the linearized gravitational equations of motion around flat space. 

More explicitly, suppose that we are only considering higher curvature theories of gravity. Then, following the arguments of  \cite{Bueno16-1}:
\beq
\begin{split}
&\frac{\kappa}{2\pi}\delta\left(S_{\text{Wald}}-\frac{K}{2G}\bar{W}\right)_{\text{higher}}=-\frac{1}{8\pi G}\eta_{bc}U_{a}U_{d}\\
&\times\left(-2\partial_{b}\partial_{c}\delta P^{abcd}_{\text{higher}}(0)\right)\left(\frac{2\Omega_{D-2}\ell^{D-1}}{(D^{2}-1)}\right)+\mathcal{O}(\ell^{D+1})\;.
\end{split}
\eeq
Meanwhile, 
\beq \delta H^{m}_{\zeta}=\delta T^{ad}U_{a}U_{d}\left(\frac{2\Omega_{D-2}\ell^{D-1}}{(D^{2}-1)}\right)+\mathcal{O}(\ell^{D+1})\;.\eeq
Therefore,
\beq
\begin{split}
&\frac{\kappa}{2\pi}\delta\left(S_{\text{Wald}}-\frac{K}{2G}\bar{W}\right)_{\text{higher}}=-\delta H^{m}_{\zeta}\\
&\Rightarrow -2\partial_{b}\partial_{c}\delta P^{abcd}_{\text{higher}}(0)=8\pi G\delta T^{ad}
\end{split}
\eeq
which exactly matches what is found in appendix C of \cite{Bueno16-1}. The Einstein contribution can be dealt with following the method described in \cite{Jacobson16-1}, and as briefly described above.

The condition (\ref{FLCDnoether}) can be understood as the Iyer-Wald identity for a theory of gravity for the geometric set-up of a causal diamond:
\beq \frac{\kappa}{2\pi}\delta \left(S_{\text{Wald}}-\frac{K}{2 G} \bar{W}\right)+ \delta H^{\zeta}_{m}=\int_{B}\delta C_{\zeta}\;,\label{1stlawcdoff}\eeq
where $\delta C_{\zeta}$ is the linearized constraint that the gravitational field equations hold. 

Following \cite{Bueno16-1} one finds that the first law of causal diamond mechanics can be understood as the Iyer-Wald identity \cite{Iyer:1994ys} in the case of a conformal Killing horizon as opposed to the dynamical horizon of a black hole. In this picture the generalized volume can be interpreted as the variation of the gravitational Hamiltonian. The first two terms on the LHS of (\ref{1stlawcdoff}), moreover, can be combined into a single object, namely, the variation of the Wald entropy keeping $\bar{W}$ held constant, i.e., 
\beq \frac{\kappa}{2\pi}\delta \left(S_{\text{Wald}}-\frac{K}{2 G}\bar{W}\right)=\frac{\kappa}{2\pi}\delta S_{\text{Wald}}|_{\bar{W}}\;,\eeq
leading to 
\beq \frac{\kappa}{2\pi}\delta S_{\text{Wald}}|_{\bar{W}}+ \delta H^{\zeta}_{m}=\int_{B}\delta C_{\zeta}\;.\label{1stlawcdoff2}\eeq

As identified in \cite{Bueno16-1}, the Wald formalism contains (JKM) ambiguities in how the Noether current and Noether charge are defined. In particular we may add an exact form $dY$ that is linear in the field variations and their derivatives to the Noether current, and $Y$ to the Noether charge. This would modify both the entropy $S_{\text{Wald}}$ and $\bar{W}$. However, as verified in \cite{Bueno16-1}, the combined modification cancel, and one may write
\beq \frac{\kappa}{2\pi}\delta S_{\text{Wald}}|_{\bar{W}}=\frac{\kappa}{2\pi}\delta(S_{\text{Wald}}+S_{JKM})|_{\bar{W}'}\;,\label{1stlawcdjkm}\eeq
where $\bar{W}'=\bar{W}+\bar{W}_{JKM}$. This shows that the resolution of the JKM ambiguity yields the same on-shell first law, provided the Wald entropy and generalized volume are modified by an exact form $dY$. 


\subsection{Entanglement Equilibrium}
\indent

Let us now show how the first law of causal diamond mechanics -- an off-shell geometric identity -- is related to a condition on entanglement. In an effective field theory the entanglement entropy can be computed using the replica trick \cite{Calabrese:2009qy}, where one defines the entropy as
\beq S_{EE}=(n\partial_{n}-1)I_{\text{eff}}(n)|_{n=1}\;,\eeq
where the effective action $I_{\text{eff}}(n)$ is evaluated on an orbifold with a conical singularity at the entangling surface with excess angle $2\pi(n-1)$. If a covariant regulator is used to define the theory, the resulting expression for the entanglement entropy is a local integral of diffeomorphism invariant contributions. When the entangling surface is the bifurcation surface of a stationary horizon, the entanglement entropy is simply the Wald entropy. In the case of nonstationary entangling surfaces, the computation can be accomplished used squashed cone techniques \cite{Fursaev:2013fta}, leading to extrinsic curvature modifications of the Wald entropy \cite{Dong:2013qoa} -- the so-called Jacobson-Myers entropy \cite{Jacobson:1993vj}. As discussed in \cite{Bueno16-1}, the extrinsic curvature modifications of the Wald entropy may be identified with the JKM ambiguities mentioned above. Thus, the entanglement entropy is given by the Wald entropy modified by specific JKM terms, i.e., the Jacobson-Myers entropy. 

This realization allows us to relate the entanglement entropy to our off-shell geometric identity (\ref{1stlawcdjkm}). The below discussion closely follows \cite{Jacobson16-1, Bueno16-1}. As briefly described in the introduction, we are performing a simultaneous geometric and quantum state variation of the entanglement entropy in a causal diamond. Therefore, the variation of the entanglement entropy $\delta S_{EE}$ includes a UV, state-independent contribution and an IR state-dependent contribution
\beq \delta S_{EE}=\delta S_{UV}+\delta S_{IR}\;.\eeq
The IR contribution describes states of a QFT in a background spacetime, while the UV contribution represents short distance physics, including quantum gravitational degrees of freedom. We should point out here that we are positing that the Hilbert space of states on $B$ can be factorized into IR and UV contributions, $\mathcal{H}_{B}=\mathcal{H}_{UV}\otimes\mathcal{H}_{IR}$, i.e., entanglement separability -- there is minimal entanglement among degrees of freedom at widely separated energy scales. 

 Upon a UV completion, the entanglement entropy in a spatial region is finite in any state, with leading term proportional to the area of the boundary of the region, and higher order contributions described by the Wald entropy. Therefore, when the geometry is varied, the entanglement entropy in the diamond (which is equivalent to entanglement in $B$) from the UV degrees of freedom near the boundary $\partial B$ will change by 
\beq \delta S_{UV}=\delta S_{\text{Wald}}^{(\epsilon)}\;.\eeq

The scale of UV completion  $\epsilon$ -- which we take to be below the Planck scale -- is such that $\mathcal{H}_{IR}$ and $\mathcal{H}_{UV}$ contain degrees of freedom with energies above and below $\epsilon$. We take the size $\ell$ of the causal diamond to be such that $L_{\text{Planck}}<\ell<1/\epsilon$. The separation between UV and IR degrees of freedom allow us to define the IR vacuum state of the ball $B$
\beq \rho_{IR}=\text{tr}_{UV}\rho\;,\eeq
where $\rho$ is the total quantum state of the diamond. Formally we may write $\rho_{IR}$ as a thermal state
\beq \rho_{IR}=\frac{1}{Z}e^{-H_{\text{mod}}}\;,\eeq
where $H_{\text{mod}}$ is the modular Hamiltonian and $Z$ is the partition function. In Minkowski space, the causal diamond may be  conformally transformed to the (planar) Rindler wedge. The Bisognano-Wichmann theorem then allows us to interpret $\rho_{IR}$ as a true thermal state with respect to the Hamiltonian generating time-translation; in the case of a conformal field theory the modular Hamiltonian will take a specific form in terms of the matter Hamiltonian $H^{m}_{\zeta}$ (\ref{matHam}) \cite{Casini:2011kv}
\beq H_{\text{mod}}=\frac{2\pi}{\kappa}H^{m}_{\zeta}\;,  \label{Hmod}\eeq
i.e., the Hamiltonian generating flow along the CKV $\zeta$. 

The entanglement entropy due to IR degrees of freedom $S_{\text{IR}}=-\text{tr}\rho_{IR}\log\rho_{IR}$ will satisfy the first law of entanglement entropy \cite{Blanco:2013joa,Wong:2013gua}
\beq \delta S_{IR}=\delta\langle H_{\text{mod}}\rangle\;.\eeq
We shall make the further conjecture, and assume that the variation of the modular Hamiltonian will carry an additional term $\delta X$ that is a spacetime scalar such that 
\beq \delta \langle H_{\text{mod}}\rangle=\frac{2\pi}{\kappa}\delta\int_{B}dB_{a}(T^{ab}\zeta_{b}+Xg^{ab}\zeta_{b})\;.\eeq
Such a conjecture was  made in \cite{Jacobson16-1}. There one assumes, to leading order that $\delta\langle H_{\text{mod}}\rangle\propto(\delta \langle T_{00}\rangle+\delta X)$, which has been shown to be a correct assumption  \cite{Casini:2016rwj,Carroll:2016lku}, though $\delta X$ may depend on $\ell$. 

Adding this to our total variation of $\delta S_{EE}$, we have a modified first law of EE
\beq \delta S_{EE}=\delta(S_{\text{Wald}}+S_{JKM})+\delta\langle H_{\text{mod}}\rangle\;. \label{1stlawEEmod}\eeq

We may now postulate the equilibrium condition: A small diamond is in equilibrium if the quantum fields are in a vacuum state and the curvature is that of a MSS, e.g., Minkowski space. Moreover, motivated by the first law of causal diamond mechanics, we require that $B$ has the same $\bar{W}'$ as in vacuum. With this, we substitute (\ref{1stlawEEmod}) into (\ref{1stlawcdjkm}), using (\ref{Hmod}), leading to
\beq \frac{\kappa}{2\pi}\delta S_{EE}|_{\bar{W}'}=\int_{B}\delta C_{\zeta}\;,\eeq
which is valid for minimally coupled, conformally invariant matter fields. 

When the variation of $\delta S_{EE}$ vanishes, we recover (\ref{lineareqnsmot}). We therefore arrive to an equivalence between the following statements: (i) the entanglement entropy $S_{EE}$ is maximal in vacuum for all (small) balls in all frames, and (ii) the linearized higher derivative equations hold everywhere. That is, the entanglement equilibrium condition is equivalent to the linearized higher derivative equations of motion to be satisfied, and vice versa. The verification of this equivalence can be found in the appendix of \cite{Bueno16-1}, which we will not repeat here but was described earlier.


\section{Failure of Killing's Identity}
\label{sec:appendB}
\indent

In our derivation of the gravitational equations of motion via the thermodynamics of causal diamonds, we made use of the conformal Killing equation
\beq \nabla_{a}\zeta_{b}+\nabla_{b}\zeta_{a}=2\Omega g_{ab}\;,\eeq
and the conformal Killing identity
\beq \nabla_{b}\nabla_{c}\zeta_{d}=R^{e}_{\;bcd}\zeta_{e}+(\nabla_{c}\Omega)g_{bd}+(\nabla_{b}\Omega)g_{cd}-(\nabla_{d}\Omega)g_{bc}\;.\eeq
An arbitrary spacetime, however, does not admit a global conformal Killing vector, therefore $\zeta^{a}$ can be understood as an \emph{approximate} conformal Killing vector. More precisely, $\zeta_{a}$ will fail to be a conformal Killing vector to some order in a Riemann normal coordinate expansion of the arbitrary spacetime (\ref{RNC}). The order at which these quantities fail depends on the order of the vector itself. The conformal Killing vector $\zeta^{a}$ we used 
\beq
\begin{split}
 \zeta^{a}&=\left(\frac{\ell^{2}-r^{2}-t^{2}}{\ell^{2}}\right)\partial^{a}_{t}-\frac{2rt}{\ell^{2}}\partial^{a}_{r}\\
& = \left(\frac{\ell^{2}-r^{2}-t^{2}}{\ell^{2}}\right)\partial^{a}_{t}-\frac{2x^{i}t}{\ell^{2}}\partial^{a}_{i}\;,
\end{split}
\eeq
with $\Omega=-2t/\ell^{2}$, was specific to $D$-dimensional Minkowski space, and is of order $\zeta^{a}=\mathcal{O}(0)+\mathcal{O}(x^{2})$, where the $\mathcal{O}(0)$ contribution is a constant. From this one finds that in an arbitrary spacetime $\zeta_{a}$ will fail the conformal Killing equation to order $\mathcal{O}(x)+\mathcal{O}(x^{3})$ and the Killing identity to order $\mathcal{O}(0)+\mathcal{O}(x^{2})$. Note that the term we keep in deriving the equations of motion, namely the integrand of\footnote{Here we ignore the vector $N_{a}$ since it will be contracted with all terms in the integrand, including the higher order contributions we neglected.}
\beq
 \int_{\Sigma}d\Sigma_{a}\left(P^{abcd}R_{ebcd}\zeta^{e}-2\zeta_{d}\nabla_{b}\nabla_{c}P^{abcd}\right)\;, \label{intdes}\eeq
is, $\mathcal{O}(0)+\mathcal{O}(x^{2})$. However, since $d\Sigma_{a}=N_{a}dAd\tau$, with $N_{a}\propto x_{i}/r$, the $\mathcal{O}(0)$ contributions vanish due to the fact we are integrating over a spherical subregion for which $\int_{\partial B}x_{i}dA=0$. Therefore, we need only concern ourselves with the $\mathcal{O}(x^{2})$ contributions coming from the failure of the conformal Killing identity. 

We realize, in fact, that the only contribution of the conformal Killing identity we made use of was the term proportional to the Riemann tensor, $R_{ebcd}\zeta^{e}$ -- we neglected all other contributions. This means that we effectively treated $\zeta^{a}$ as an approximate Killing vector rather than an approximate conformal Killing vector. We therefore find ourselves in a similar situation as the authors of \cite{Parikh:2017aas}: We must remove the higher order contributions coming from the failure of Killing's identity. Specifically, in the integrand (\ref{intdes}), the term $P^{abcd}\nabla_{b}\nabla_{c}\zeta_{d}$ should be replaced with
 \beq P^{abcd}\nabla_{b}\nabla_{c}\zeta_{d}=P^{abcd}R_{ebcd}\zeta^{e}+P^{abcd}f_{bcd}\;,\eeq
with
\beq f_{bcd}=\nabla_{b}\nabla_{c}\zeta_{d}-R_{ebcd}\zeta^{e}-(\nabla_{c}\Omega)g_{bd}+(\nabla_{d}\Omega)g_{bc}\;,\eeq
from which we see that $f_{bdc}=-f_{bcd}$. Here $f_{bcd}$ quantifies the failure of Killing's identity. Our task is therefore to find a way to eliminate
\beq \int_{\Sigma}d\Sigma_{a}P^{abcd}f_{bcd}\;,\label{undesired}\eeq
at least to the order at which we keep the desired contribution $\int_{\Sigma} d\Sigma_{a}P^{abcd}R_{ebcd}\zeta^{e}$. Specifically the integrand we wish to keep
\beq N_{a}P^{abcd}R_{bcde}\zeta^{e}\;,\eeq
goes like $\mathcal{O}(0)+\mathcal{O}(x^{2})$. The $\mathcal{O}(0)$ contribution, as mentioned above, vanishes due to the fact we are integrating over a spherical subregion. Therefore, the order of the integrand we are interested in keeping is $\mathcal{O}(x^{2})$, and we must remove the $\mathcal{O}(x^{2})$ contributions of the undesired term. 

To study this problem we introduce the notation
\beq f_{bcd}=f^{(0)}_{bcd}+f_{bcd}^{(1)}+f_{bcd}^{(2)}+...\;,\eeq
where $f^{(0)}_{bcd}$ denotes the $\mathcal{O}(0)$ contribution to $f_{bcd}$, $f^{(1)}_{bcd}$ the $\mathcal{O}(x)$ contribution, and so forth. We will use this notation to decompose each object appearing in the integrand (\ref{undesired}), i.e., $N_{a}=N_{a}^{(0)}$, and $P^{abcd}=P^{abcd}_{(0)}+P^{abcd}_{(1)}+...$. 

In order to remove contribution (\ref{undesired}) to the desired order, we will follow the method developed in \cite{Parikh:2017aas}, by modifying $\zeta_{a}$ and $N_{a}$, by adding undetermined higher order contributions to $\zeta_{a}$. The algorithm for removing the terms can be described as follows: The integrand of (\ref{undesired}) is a collection of monomials. Because we are integrating over a spherical subregion, many of these monomial contributions will vanish, e.g., when the integrand goes like $tx_{i}/r$. Some terms will remain, however, and the only way to remove these contributions is to add in higher order modifications to $\zeta^{a}$, e.g., 
\beq \zeta_{a}=\left(\frac{\ell^{2}-r^{2}-t^{2}}{\ell^{2}}\right)\partial^{a}_{t}-\frac{2x^{i}t}{\ell^{2}}\partial^{a}_{i}+\frac{1}{3!}D_{a\mu\nu\rho}x^{\mu}x^{\nu}x^{\rho}+...\;,\eeq
where here the greek indices $\mu,\nu$ run over the whole spacetime index. We can likewise modify $N_{a}$. These modifications to $\zeta_{a}$ will include additional contributions to $f_{bcd}$ of the same monomial structure as before. We then choose the undetermined coefficients $D_{a\mu\nu\rho}$, etc. so as to cancel these terms. In essence we add counterterms to $\zeta_{a}$ to remove (\ref{undesired}) to the desired order. One problem which may arise is whether there are enough undetermined coefficients to cancel all of the monomials which may appear. 

Putting all of this together, the lowest order contribution in the integrand of the offending term (\ref{undesired}) is
\beq \int_{\Sigma}dAd\tau n_{a}^{(0)}P^{abcd}_{(0)}f^{(0)}_{bcd}\;.\eeq
As already discussed, this term vanishes via parity arguments. The next order term in the integrand is $\mathcal{O}(x)$,
\beq \int_{\Sigma}dAd\tau\biggr\{N^{(1)}_{a}P^{abcd}_{(0)}f^{(0)}_{bcd}+N^{(0)}_{a}P^{abcd}_{(1)}f^{(0)}_{bcd}+N^{(0)}_{a}P^{abcd}_{(0)}f^{(1)}_{bcd}\biggr\}\;, \label{NPfO1}\eeq
and the $\mathcal{O}(x^{2})$ term we must remove is
\begin{widetext}
\beq 
\begin{split}
&\int_{\Sigma}dAd\tau\biggr\{N^{(0)}_{a}P^{abcd}_{(2)}f_{bcd}^{(0)}+N^{(0)}_{a}P^{abcd}_{(0)}f_{bcd}^{(2)}+N^{(0)}_{a}P^{abcd}_{(1)}f_{bcd}^{(1)}+N_{a}^{(1)}P^{abcd}_{(0)}f^{(1)}_{bcd}+N^{(1)}_{a}P^{abcd}_{(1)}f_{bcd}^{(0)}\\
&+N^{(2)}_{a}P^{abcd}_{(0)}f^{(0)}_{bcd}+\sqrt{h}N_{a}^{(0)}P^{abcd}_{(0)}f_{bcd}^{(0)}\biggr\}\;.
\end{split}
\label{NPfO2}\eeq
\end{widetext}
As we will see, we can in fact drop the terms proportional to $N^{(1)}_{a}$, as it generates more work for us than it helps. 

To summarize the algorithm, in order to say we have achieved in deriving the nonlinear equations of motion for higher derivative gravity, we must show how to eliminate the above two contributions (\ref{NPfO1}) and (\ref{NPfO2}). We do this by modifying the $\zeta$ to include higher order contributions, and count the number of undetermined coefficients to see if we have enough terms to eliminate (\ref{NPfO1}) and (\ref{NPfO2}). At first glance it seems as though this is indeed possible simply by a naive counting of the number of monomials which appear in the integrand, compared to a naive counting of the number of undetermined coefficients that are available. 

\subsection{Removing $\mathcal{O}(x)$ Contributions}
\indent

First we write $f_{bcd}$ in a more useful form
\begin{widetext}
\beq 
\begin{split}
f_{bcd}&=\nabla_{b}\nabla_{c}\zeta_{d}-R_{ebcd}\zeta^{e}-(\nabla_{c}\Omega) g_{bd}+(\nabla_{d}\Omega)g_{bc}\\
&=\partial_{b}\partial_{c}\zeta_{d}+\left(2\Gamma^{f}_{\;b(c}\Gamma^{e}_{\;d)f}-\partial_{b}\Gamma^{e}_{\;cd}\right)\zeta_{e}-\left(\Gamma^{e}_{\;bc}\partial_{e}\zeta_{d}+2\Gamma^{e}_{\;d(c}\partial_{b)}\zeta_{e}\right)-R^{e}_{\;bcd}\zeta_{e}-(\nabla_{c}\Omega) g_{bd}+(\nabla_{d}\Omega)g_{bc}\;.
\end{split}
\eeq
We can drop the whole second term because it is symmetric in indices $cd$ and is being contracted with $P^{abcd}$. What remains is:
\beq
\begin{split}
 f_{bcd}&=\partial_{b}\partial_{c}\zeta_{d}-\left(\Gamma^{e}_{\;bc}\partial_{e}\zeta_{d}+2\Gamma^{e}_{\;d(c}\partial_{b)}\zeta_{e}\right)-R^{e}_{\;bcd}\zeta_{e}-(\nabla_{c}\Omega) g_{bd}+(\nabla_{d}\Omega)g_{bc}\;.
\end{split}
\eeq
We think about modifying $\zeta_{a}$ in the following way:
\beq
\begin{split}
 \zeta_{a}&=\zeta_{a}^{(0)}+\zeta_{a}^{(2)}+\zeta^{(3)}_{a}+\zeta^{(4)}_{a}+...\\
&=-\frac{1}{\ell^{2}}(\ell^{2}-r^{2}-t^{2})\partial_{a}^{t}-\frac{2tx_{i}}{\ell^{2}}\partial_{a}^{i}+\zeta^{(3)}_{a}+\zeta^{(4)}_{a}+...\;,
\end{split}
\eeq
where the $\zeta_{a}^{(0)}$ contribution is constant. A similar expansion holds for $N^{a}$. 

Let's now classify $f_{bcd}^{(0)}$. Clearly we get a contribution from $\partial_{b}\partial_{c}\zeta_{d}$, and from the $\nabla\Omega$ terms. Specifically, 
\beq
\begin{split}
 f_{bcd}^{(0)}&=\partial_{b}\partial_{c}\zeta_{d}^{(2)}-(\nabla_{c}\Omega) \eta_{bd}+(\nabla_{d}\Omega)\eta_{bc}\\
&=\partial_{b}\partial_{c}\zeta_{d}^{(2)}-\frac{2}{\ell^{2}}(\delta^{t}_{\;d}\eta_{bc}-\delta^{t}_{\;c}\eta_{bd})\;.
\end{split}
\eeq

Let's look at the $\mathcal{O}(x)$ contribution of  which would be present in (\ref{NPfO1}) even without modifying $\zeta_{a}$ or $N_{a}$. This is:
\beq 
\begin{split}
N^{(0)}_{a}P^{abcd}_{(1)}f^{(0)}_{bcd}&=N^{(0)}_{i}P^{ibcd}_{(1)}f^{(0)}_{bcd}=N^{(0)}_{i}P^{itcd}_{(1)}f^{(0)}_{tcd}+N^{(0)}_{i}P^{ijcd}_{(1)}f^{(0)}_{jcd}\\
&=N^{(0)}_{i}P^{itjd}_{(1)}f^{(0)}_{tjd}+N^{(0)}_{i}P^{ittd}_{(1)}f^{(0)}_{ttd}+N^{(0)}_{i}P^{ijtd}_{(1)}f^{(0)}_{jtd}+N^{(0)}_{i}P^{ijkd}_{(1)}f^{(0)}_{jkd}\\
&=N^{(0)}_{i}P^{itjt}_{(1)}f^{(0)}_{tjt}+N^{(0)}_{i}P^{itjk}_{(1)}f^{(0)}_{tjk}+N^{(0)}_{i}P^{ittj}_{(1)}f^{(0)}_{ttj}+N^{(0)}_{i}P^{ijtk}_{(1)}f^{(0)}_{jtk}+N^{(0)}_{i}P^{ijkt}_{(1)}f^{(0)}_{jkt}+N^{(0)}_{i}P^{ijk\ell}_{(1)}f^{(0)}_{jk\ell}\;.
\end{split}
\eeq
Thus our task is to compute 
\beq f^{(0)}_{tjt},\quad f^{(0)}_{tjk},\quad f^{(0)}_{ttj},\quad f^{(0)}_{jtk},\quad f^{(0)}_{jkt},\quad f^{(0)}_{jk\ell}\;.\eeq
It is straightforward to work out that the only non-zero term is
\beq 
\begin{split}
f^{(0)}_{tjk}&=\partial_{t}\partial_{j}\zeta_{k}-\frac{2}{\ell^{2}}(\delta^{t}_{\;d}\eta_{bc}-\delta^{t}_{\;c}\eta_{bd})|_{b=t,c=j,d=k}\\
&=-\frac{2}{\ell^{2}}\delta_{jk}+0=-\frac{2}{\ell^{2}}\delta_{jk}\;,
\end{split}
\eeq
Therefore, the only non-zero contribution will be:
\beq N_{i}^{(0)}P^{itjk}_{(1)}f^{(0)}_{tjk}\;.\eeq
But this term vanishes because $f^{(0)}_{tjk}$ is symmetric in $jk$ indices, while $P_{(1)}^{itjk}$ is antisymmetric. Thus, the entire contribution:
\beq N_{a}^{(0)}P^{abcd}_{(1)}f^{(0)}_{bcd}=0\;.\eeq
In fact, whenever we have something of the form $N^{(a)}_{(0)}P^{abcd}f^{(0)}_{bcd}$, we see that it vanishes, as we never specified the form of $P^{abcd}$ above. We will therefore be able to drop some terms appearing in the $\mathcal{O}(x^{2})$ contribution (\ref{NPfO2}) as well.

There is another term in (\ref{NPfO1}) which appears due to $\zeta_{a}$ being an approximate (conformal) Killing vector, namely, the one proportional to $f^{(1)}_{bcd}$. Without modifying $\zeta_{a}$, the only contribution to this comes from
\beq (\nabla_{d}\Omega)g_{bc}-(\nabla_{c}\Omega)g_{bd}-R^{e}_{\;bcd}\zeta^{(0)}_{e}\;.\eeq
To leading order, we have $\nabla\Omega g\sim (\nabla\Omega)(p)_{\mu}x^{\mu}\eta$, where $\eta$ is the Minkowski metric. Calling $(\nabla_{d}\Omega)_{\mu}(p)\equiv \Omega_{d\mu}(p)$, and noting that $\zeta^{(0)e}=\delta^{te}$, we find that, without modifying $\zeta_{a}$, we have: 
\beq f^{(1)}_{bcd}=(\Omega_{d\mu}x^{\mu}\eta_{bc}-\Omega_{c\mu}\eta_{bd}x^{\mu})-(R_{tbcd})_{\mu}x^{\mu}\;,\eeq
where it is understood that $(R_{tbcd})_{\mu}$ is evaluted at the point $p$. Now we work to see which of
\beq 
\begin{split}
N^{(0)}_{a}P^{abcd}_{(1)}f^{(0)}_{bcd}&=N^{(0)}_{i}P^{itjt}_{(0)}f^{(1)}_{tjt}+N^{(0)}_{i}P^{itjk}_{(0)}f^{(1)}_{tjk}+N^{(0)}_{i}P^{ittj}_{(0)}f^{(1)}_{ttj}+N^{(0)}_{i}P^{ijtk}_{(0)}f^{(1)}_{jtk}+N^{(0)}_{i}P^{ijkt}_{(0)}f^{(1)}_{jkt}\\
&+N^{(0)}_{i}P^{ijk\ell}_{(0)}f^{(1)}_{jk\ell}\;,
\end{split}
\eeq
must be cancelled. Let's work out each of the $f^{(1)}_{bcd}$. The only non-zero contributions we have include:
\beq f^{(1)}_{tjt}=\Omega_{j\mu}x^{\mu}=-f^{(1)}_{ttj}\;,\eeq
\beq f^{(1)}_{jkt}=\Omega_{t\mu}x^{\mu}\eta_{jk}-(R_{tjkt})_{\mu}x^{\mu}=-f^{(1)}_{jtk}\;,\eeq
\beq f^{(1)}_{jk\ell}=(\Omega_{\ell\mu}\eta_{jk}-\Omega_{k\mu}\eta_{j\ell})x^{\mu}-(R_{tjk\ell})_{\mu}x^{\mu}\;,\eeq
Then, using the symmetries of $P^{abcd}$ and $f^{(1)}_{bcd}$, we have:
\beq 
\begin{split}
N_{a}^{(0)}P^{abcd}_{(0)}f^{(1)}_{bcd}&=N^{(0)}_{i}P^{itjt}_{(0)}(2\Omega_{j\mu}x^{\mu})+N^{(0)}_{i}P^{ijkt}_{(0)}(2\Omega_{t\mu}x^{\mu}\eta_{jk}-2(R_{tjkt})_{\mu}x^{\mu})\\
&+N^{(0)}_{i}P^{ijk\ell}_{(0)}\left[(\Omega_{\ell\mu}\eta_{jk}-\Omega_{k\mu}\eta_{j\ell})x^{\mu}-(R_{tjk\ell})_{\mu}x^{\mu}\right]\;.
\end{split}
\eeq
\end{widetext}
Using spherical symmetry, and that $N^{(0)}_{i}=x_{i}/r$, we see that the only non-vanishing contributions to this will be when $\mu =m$ -- a spatial index, i.e., 
\beq 
\begin{split}
&\int_{\Sigma}dAd\tau\biggr\{2P^{itjt}_{(0)}(\Omega_{jm})+2P^{ijkt}_{(0)}(\Omega_{tm}\eta_{jk}-(R_{tjkt})_{m})\\
&+P^{ijk\ell}_{(0)}(\Omega_{\ell m}\eta_{jk}-\Omega_{k m}\eta_{j\ell}-(R_{tjk\ell})_{m})\biggr\}N^{(0)}_{i}x^{m}\\
&\equiv\int_{\Sigma}dAd\tau\mathcal{M}^{i}_{\;m}N^{(0)}_{i}x^{m}\;,
\end{split}
\label{N0P0f1}\eeq
where
\beq 
\begin{split}
\mathcal{M}^{i}_{\;m}&\equiv \biggr\{2P^{itjt}_{(0)}(\Omega_{jm})+2P^{ijkt}_{(0)}(\Omega_{tm}\eta_{jk}-(R_{tjkt})_{m})\\
&+P^{ijk\ell}_{(0)}(\Omega_{\ell m}\eta_{jk}-\Omega_{k m}\eta_{j\ell}-(R_{tjk\ell})_{m})\biggr\}\;.
\end{split}
\eeq
More precisely, the only non-vanishing contribution occurs when $i=m$, i.e., 
\beq \int_{\Sigma}dAd\tau\sum_{i}\mathcal{M}_{ii}\frac{(x^{i})^{2}}{r}\;.\eeq
We see then that the only type of polynomial we see appearing includes $(x_{i})^{2}/r$ -- or $(D-1)$ such terms for a $D$-dimensional spacetime.

This shows us that we must modify $\zeta_{a}$ such that we can eliminate such contributions. Consider, then, the modification
\beq \zeta^{(3)}_{d}=\frac{1}{3!}C_{\mu\nu\rho d}x^{\mu}x^{\nu}x^{\rho}\;,\eeq
where $C_{\mu\nu\rho d}$ is a collection of $D^{4}$ completely undetermined coefficients. It is easy to see that this will provide a contribution to $f^{(1)}_{bcd}$ only through
\beq \partial_{b}\partial_{c}\zeta^{(3)}_{d}=C_{\mu bcd}x^{\mu}\;.\eeq

Putting this into the integrand (\ref{N0P0f1}) we have
\beq \int_{\Sigma}dAd\tau(\mathcal{M}^{i}_{\;m}+P^{ibcd}_{(0)}C_{mbcd})N^{(0)}_{i}x^{m}\;.\eeq
Or, using spherical symmetry,
\beq \int_{\Sigma}dAd\tau\sum_{i}(\mathcal{M}_{ii}+P_{i,(0)}^{\;\;bcd}C_{ibcd})\frac{(x^{i})^{2}}{r}\;.\eeq
We see then that there are more than enough $C$ coefficients to eliminate the undesired terms. 

The only other contribution in (\ref{NPfO1}) is one which arises form the $N_{a}^{(1)}$ modification. Clearly, this term is unnecessary, and therefore we simply do not modify $N$ at this level. This then takes care of the (\ref{NPfO1}) term -- by modifying $\zeta_{a}$ at $\mathcal{O}(x^{3})$ as shown above, we can remove the undesired (\ref{NPfO1}). Let's  move on to the $\mathcal{O}(x^{2})$ contribution, (\ref{NPfO2}). 

\hspace{2mm}

\subsection{Removing $\mathcal{O}(x^{2})$ Contributions}
\indent

We first point out some simplifications we can make to (\ref{NPfO2}). Using that $N^{(0)}_{a}P^{abcd}f^{(0)}_{bcd}$ all cancel, we can neglect all such terms. Likewise, we can drop any term proportional to $N^{(1)}_{a}$. Thus, we have
\beq 
\begin{split}
&\int_{\Sigma}dAd\tau\biggr\{n^{(0)}_{a}P^{abcd}_{(0)}f_{bcd}^{(2)}+n^{(0)}_{a}P^{abcd}_{(1)}f_{bcd}^{(1)}+n^{(2)}_{a}P^{abcd}_{(0)}f^{(0)}_{bcd}\biggr\}\;.
\end{split}
\label{NPfO2re}\eeq
A priori we have no reason to drop the $N_{a}^{(2)}$ modification, however, as we will see, we may drop it simply because we have enough coefficients to eliminate all undesired terms, leaving us with two terms. Note that $N^{(0)}_{a}P^{abcd}_{(1)}f^{(1)}_{bcd}$ will include contributions both from the failure of $\zeta$ being a Killing vector, and from us modifying $\zeta_{a}$. This means we bring in a large number of $C$ coefficients, potentially all $D^{4}$ of them. However, $(D-1)$ of these coefficients we potentially used, while many others cannot be used due to the fact we are integrating over a co-dimension-2 sphere. Thus, while there are a handful of remaining $C$ coefficients which can be used to eliminate the $\mathcal{O}(x^{2})$ integrand, we cannot rely on or assume we have each coefficient; we must look to modifying $\zeta_{a}$ by adding a term of the form
\beq \zeta^{(4)}_{a}=\frac{1}{4!}D_{\mu\nu\rho\sigma a}x^{\mu}x^{\nu}x^{\sigma}x^{\rho}\;,\eeq
which we see has $D^{5}$ undetermined coefficients. Therefore, by a naive counting argument we find that we will have more than enough $D$ and remaining $C$ coefficients to eliminate all undesired contributions at the $\mathcal{O}(x^{2})$ level.

Begin with
\beq 
\begin{split}
&N^{(0)}_{a}P^{abcd}_{(0)}f^{(2)}_{bcd}=N^{(0)}_{i}P^{itjt}_{(0)}(f^{(2)}_{tjt}-f^{(2)}_{ttj})\\
&+N^{(0)}_{i}P^{itjk}_{(0)}f^{(2)}_{tjk}+N^{(0)}_{i}P^{ijtk}_{(0)}(f^{(2)}_{jtk}-f^{(2)}_{jkt})+N^{(0)}_{i}P^{ijk\ell}_{(0)}f^{(2)}_{jk\ell}\;,
\end{split}
\eeq
where
\begin{widetext}
\beq
\begin{split}
 f^{(2)}_{bcd}&=\partial_{b}\partial_{c}\zeta^{(4)}_{d}-(\Gamma^{e}_{\;bc}\partial_{e}\zeta^{(2)}_{d}+2\Gamma^{e}_{\;d(c}\partial_{b)}\zeta^{(2)}_{e})-R^{e}_{\;bcd}(p)\zeta^{(2)}_{e}-(\nabla_{c}\Omega)h_{bd}+(\nabla_{d}\Omega)h_{bc}\\
&-\frac{1}{2}(\nabla_{c}\Omega)_{\mu\nu}x^{\mu}x^{\nu}\eta_{bd}+\frac{1}{2}(\nabla_{d}\Omega)_{\mu\nu}x^{\mu}x^{\nu}\eta_{bc}\;,
\end{split}
\eeq
with
\beq h_{bd}=-\frac{1}{3}R_{b\mu d\nu}(p)x^{\mu}x^{\nu}\;.\eeq
Following a similar strategy to remove $\mathcal{O}(x)$ contributions and using \cite{Parikh:2017aas} as a guide, several lines of algebra later show that
\beq 
\begin{split}
N_{a}^{(0)}P^{abcd}_{(0)}f^{(2)}_{bcd}&=N^{(0)}_{i}P^{itjt}_{(0)}\left[\left(\frac{1}{2}(D_{\mu\nu tjt}-D_{\mu\nu ttj})+\frac{4}{3\ell^{2}}R_{t\mu j\nu}(p)+\Omega_{j\mu\nu}\right)x^{\mu}x^{\nu}+\frac{4}{\ell^{2}}R_{ktjt}(p)tx^{k}\right]\\
&+N^{(0)}_{i}P^{itjk}_{(0)}\left[\frac{1}{2}D_{\mu\nu tjk}x^{\mu}x^{\nu}+\frac{2}{\ell^{2}}R_{\ell tjk}(p)tx^{\ell}\right]\\
&+N^{(0)}_{i}P^{ijtk}_{(0)}\left[\left(\frac{1}{2}(D_{\mu\nu jtk}-D_{\mu\nu jkt})-\frac{4}{3\ell^{2}}R_{j\mu k\nu}(p)-\Omega_{t\mu\nu}\delta_{jk}\right)x^{\mu}x^{\nu}+\frac{4}{\ell^{2}}R_{\ell jtk}(p)tx^{\ell}\right]\\
&+N^{(0)}_{i}P^{ijk\ell}_{(0)}\left[\left(\frac{1}{2}D_{\mu\nu jk\ell}+\frac{1}{2}(\Omega_{\ell\mu\nu}\delta_{jk}-\Omega_{k\mu\nu}\delta_{j\ell})\right)x^{\mu}x^{\nu}+\frac{2}{\ell^{2}}R_{mjk\ell}(p)tx^{m}\right]\;.
\end{split}
\label{n0P0f2}\;,\eeq
and
\beq 
\begin{split}
&N_{a}^{(0)}P^{abcd}_{(1)}f^{(1)}_{bcd}\\
&=\biggr\{(P^{ibcd}_{(1)})_{\nu}C_{\mu bcd}+(P^{itjt}_{(1)})_{\nu}(2\Omega_{j\mu})+(P^{ijkt}_{(1)})_{\nu}(2\Omega_{t\mu}\delta_{jk}-2(R_{tjkt})_{\mu})+(P^{ijk\ell}_{(1)})_{\nu}\left[(\Omega_{\ell\mu}\eta_{jk}-\Omega_{k\mu}\eta_{j\ell})-(R_{tjk\ell})_{\mu}\right]\biggr\}N^{(0)}_{i}x^{\mu}x^{\nu}\;.
\end{split}
\label{n0P1f1}\;,\eeq
\end{widetext}
where we have written $P^{abcd}_{(1)}(x)=(P^{abcd}_{(1)})_{\nu}x^{\nu}$. Since $N_{i}^{(0)}\propto x_{i}$, this fixes what $\mu,\nu$ have to be. Either $\mu =0,\nu=j=i$ or $\mu=j=i,\nu=0$. All other contributions vanish due to integration. 

We would now add together (\ref{n0P0f2}) and (\ref{n0P1f1}) in the integrand (\ref{NPfO2re}). We see that we have enough $D$ coefficients to cancel these terms, without introducing $N^{(2)}_{a}$. This can be explicitly checked in the case of $f(R)$ gravity in $2+1$ dimensions -- the most restrictive example. Since we have more than enough coefficients to account for the above monomial contributions, we need not modify $N_{a}$ at all, and may therefore have eliminated (\ref{NPfO2re}). This completes the derivation of the equations of motion.

\section{Iyer-Wald Identity For Stretched Lightcones}
\label{sec:appendC}
\indent

Here, after reviewing the basic set-up of the Iyer-Wald formalism \cite{Iyer:1994ys}, we consider the Iyer-Wald identity for the geometry of future stretched lightcones. We will closely follow the arguments presented in \cite{Bueno16-1} due to the geometric similarity between the stretched lightcone and causal diamond. 

\subsection{Iyer-Wald Formalism}
\indent

Let $L[\phi]$ be the local spacetime $D$-form Lagrangian of a general diffeomorphism invariant theory, where $\phi$ represents a collection of dynamical fields, e.g., the metric and matter fields. Varying the Lagrangian yields
\beq \delta L=E\cdot\delta\phi+d\theta[\delta\phi]\;,\eeq
where $E$ denotes the equations of moton for all of the dynamical fields, and $\theta$ is the symplectic potential $(D-1)$-form. The antisymmetric variation of $\theta$ leads to the symplectic current, a $(D-1)$-form,
\beq \omega[\delta_{1}\phi,\delta_{2}\phi]=\delta_{1}\theta[\delta_{2}\phi]-\delta_{2}\theta[\delta_{1}\phi]\;,\label{sympcurrent}\eeq
whose integral over a Cauchy surface $B$ gives the symplectic form for the phase description of the theory. Given an arbitrary vector field $\xi^{a}$, evaluating the symplectic form on the Lie derivative $\mathcal{L}_{\xi}\phi$ yields the variation of the Hamiltonian $H_{\xi}$ which generates the flow $\xi^{a}$:
\beq\delta H_{\xi}=\int_{B}\omega[\delta\phi,\mathcal{L}_{\xi}\phi]\;. \label{Hamilxi}\eeq
Now take $B$ to be a ball-shaped region, and let $\xi^{a}$ be a future-pointed, timelike vector that vanishes on the boundary $\partial B$. When the background geometry satisfies the field equations $E=0$, , and $\xi$ vanishes on $\partial B$, we arrive to  Wald's variational identity 
\beq \int_{B}\omega[\delta\phi,\mathcal{L}_{\xi}\phi]=\int_{B}\delta J_{\xi}\;,\label{Waldvarid}\eeq
where we have introduced the Noether current $J_{\xi}$
\beq J_{\xi}=\theta[\mathcal{L}_{\xi}\phi]-i_{\xi}L\;,\eeq
with $i_{\xi}$ representing the contraction of the vector $\xi^{a}$ on the first index of the differential form. Recall that the Noether current $J_{\xi}$ can always be written as \cite{Iyer:1995kg}
\beq J_{\xi}=dQ_{\xi}+C_{\xi}\;,\label{Noethercurrent}\eeq
where $Q_{\xi}$ is the Noether charge $(D-2)$-form and $C_{\xi}$ are the constraint field equations associated with diffeomorphism gauge symmetry. When we assume that the matter equations are imposed, one finds 
\beq C_{\xi}=-2\xi^{a}E_{a}^{\;b}\epsilon_{b}\;,\eeq
where $E^{ab}$ is the variation of the Lagrangian density with respect to the metric, and $\epsilon_{a}$ is the volume form on $B$. Combining (\ref{Hamilxi}), (\ref{Waldvarid}), and (\ref{Noethercurrent}) leads to the Iyer-Wald identity:
\beq -\int_{\partial B}\delta Q_{\xi}+\delta H_{\xi}=\int_{B}\delta C_{\xi}\;.\label{IWid}\eeq
When the linearized constraints hold, $\delta C_{\xi}=0$, the variation of the Hamiltonian is a boundary integral of $\delta Q_{\xi}$. We will show that this off-shell identity leads to the first law of stretched lightcones. Observe that, unlike the case with black hole thermodynamics, $\delta H_{\xi}$ here is non-vanishing; this is because $\xi^{a}$ is not a true Killing vector. 

Let us proceed and evaluate the Iyer-Wald identity (\ref{IWid}) for an arbitrary theory of gravity for the geometric set-up for the stretched lightcone described above (\ref{sec:geosetup}). Here we will make the simplifying assumption that the matter fields are minimally coupled, such that the Lagrangian splits into metric and matter contributions
\beq L=L^{g}+L^{m}\;,\eeq
with $L^{g}$ being an arbitrary diffeomorphism-invariant function of the metric, Riemann tensor, and the covariant derivatives of the Riemann tensor\footnote{In our discussion above we did not consider theories of gravity which also depend on derivatives of the Riemann tensor, however, it is easy to modify our arguments to include such theories -- in the case one perturbs around maximally symmetric spacetimes.}. This separation allows us to also decompose the symplectic potential and the Hamiltonian as $\theta=\theta^{g}+\theta^{m}$, and $\delta H_{\xi}=\delta H^{g}_{\xi}+\delta H^{m}_{\xi}$. Therefore, the Iyer-Wald identity (\ref{IWid}) becomes
\beq -\int_{\partial B}\delta Q_{\xi}+\delta H^{g}_{\xi}+\delta H^{m}_{\xi}=\int_{B}\delta C_{\xi}\;.\eeq

We can relate the integrated Noether charge to the Wald entropy via \cite{Wald93-1}:
\beq -\int_{\partial B}Q_{\xi}= 4GS_{\text{Wald}}\;.\eeq
where $G$ is Newton's gravitational constant, and the Wald entropy functional $S_{\text{Wald}}$ is
\beq S_{\text{Wald}}=-\frac{1}{4G}\int_{\partial B}dS_{ab}(P^{abcd}\nabla_{c}\xi_{d}-2\xi_{d}\nabla_{c}P^{abcd})\;,\eeq
with $dS_{ab}=\frac{1}{2}(n_{a}u_{b}-n_{b}u_{a})dA$\footnote{A brief comment on notation: For comparison to \cite{Bueno16-1}, we note that there the authors choose the convention where $1/4G \to 2\pi$, and use that the Wald entropy is written as
\beq S_{\text{Wald}}=-2\pi\int_{\partial B}\mu P^{abcd}n_{ab}n_{cd}\;,\eeq
where $\mu$ is the volume form on $\partial B$, which $\epsilon_{ab}=-n_{ab}\wedge\mu$.}. Following, \cite{Iyer:1994ys}, this relationship also holds for first order perturbations
\beq \int_{\partial B}\delta Q_{\xi}=-4G\delta S_{\text{Wald}}\;.\eeq

Our next task is to evaluate the variation of the gravitational Hamiltonian $\delta H_{\xi}^{g}$. As we detail below, this leads us to the derivation of the generalized area of stretched lightcones, analogous to the generalized volume of causal diamonds constructed in \cite{Bueno16-1}. 

\subsection{Generalized Area of Stretched Lightcones}
\indent

Here we closely follow the arguments presented in \cite{Bueno16-1} to work out the variation of the gravitational Hamiltonian for an arbitrary theory of gravity in the geometric set-up of the stretched lightcone. In the calculation that follows we will consider the case of looking at perturbations about a maximally symmetric background (MSS), specifically Minkowski space. Along the way we will mention how some of these assumptions might be relaxed. 

For a Lagrangian that depends on the Riemann tensor and its covariant derivatives, the symplectic potential $\theta^{g}$ is given by 
\beq \theta^{g}=2P^{bcd}\nabla_{d}\delta g_{bc}+S^{ab}\delta g_{ab}+\sum_{i=1}^{m-1}T_{i}^{abcda_{1}...a_{i}}\delta\nabla_{(a_{1}}...\nabla_{a_{i})}R_{abcd}\;,\eeq
where we use the notation of \cite{Bueno16-1} such that $P^{bcd}=\epsilon_{a}P^{abcd}$, and $S^{ab}$ and $T_{i}^{abcd...}$ are locally constructed from the metric, its curvature, and covariant derivatives of the curvature. Due to the antisymmetry of $P^{bcd}$ in $c$ and $d$, the symplectic current (\ref{sympcurrent}) takes the form
\beq 
\begin{split}
\omega^{g}&=2\delta_{1}E^{bcd}\nabla_{d}\delta_{2}g_{bc}-2E^{bcd}\delta_{1}\Gamma^{e}_{\;db}\delta_{2}g_{ec}+\delta_{1} S^{ab}\delta_{2}g_{ab}\\
&+\sum_{i=1}^{m-1}\delta_{1}T_{i}^{abcda_{1}...a_{i}}\delta_{2}\nabla_{(a_{1}}...\nabla_{a_{i})}R_{abcd}-(1\leftrightarrow2)\;. 
\end{split}
\label{sympcurrent2}\eeq

Let's now employ the geometric set-up discussed above. We use the fact that we are perturbing around a maximally symmetric background. This allows us to write
\beq R_{abcd}=\frac{R}{D(D-1)}(g_{ac}g_{bd}-g_{ad}g_{bc})\;,\eeq
with a constant Ricci scalar $R$, such that 
\beq \nabla_{e}R_{abcd}=0\;,\quad\mathcal{L}_{\xi}R_{abcd}|_{t=0}=0\;.\eeq
Moreover, since the tensors $P^{abcd}$, $S^{ab}$ and $T_{i}^{abcd...}$ are all constructed from the metric and curvature, they will also have vanishing Lie derivatives along $\xi^{a}$, when evaluated on $B$. 

If we replace $\delta_{2}g_{ab}$ in (\ref{sympcurrent2}) with $\mathcal{L}_{\xi}g_{ab}$, and make use of (\ref{nabLg})
\beq \nabla_{d}(\mathcal{L}_{\xi}g_{ab})|_{t=0}=\frac{2}{N_{\xi}}u_{d}\tilde{g}_{ab}\;,\eeq
with
\beq \tilde{g}_{ab}=\delta_{a}^{i}\delta^{j}_{b}\left(\delta_{ij}-\frac{x_{i}x_{j}}{r^{2}}\right)\;,\eeq
then,
\beq 
\begin{split}
&\omega^{g}[\delta g,\mathcal{L}_{\xi}g]|_{B}=\frac{2}{N_{\xi}}\biggr\{2\tilde{g}_{bc}u_{d}\delta P^{bcd}+P^{bcd}\{u_{b}\tilde{\delta}^{e}_{\;d}\delta g_{ce}\\
&+u_{d}\tilde{\delta}^{e}_{\;b}\delta g_{ce}-u^{e}\tilde{g}_{db}\delta g_{ce}\}\biggr\}\;.
\end{split}
\eeq

Following similar computations performed in \cite{Bueno16-1} we find to leading order in the RNC
\beq \omega[\delta g,\mathcal{L}_{g}]|_{B}=-\delta[\frac{4}{N}\eta P^{abcd}U_{a}u_{d}\tilde{g}_{bc}]\;.\eeq
Showing this takes quite a few lines of algebra, however, when all is said and done, we can take (33) of \cite{Bueno16-1} and simply replace $g_{bc}$ with $\tilde{g}_{bc}$. 

Thus, we are varying the object
\beq \int_{B}dB_{a}\frac{\alpha}{r^{2}}P^{abcd}u_{d}\tilde{g}_{bc}\;.\eeq
However, after converting back to the conventions used in the body of this paper, we find that 
\beq\delta H_{\xi}^{g}=-\frac{1}{2\pi\alpha}\delta \tilde{S}\;,\eeq
 i.e., the entropy due to the natural expansion of the hyperboloid $\bar{S}$ (\ref{genarea}). 

In summary, we have arrived to the off-shell variational identity 
\beq \frac{1}{2\pi\alpha}\delta (S_{\text{Wald}}-\bar{S})+\delta H^{m}_{\xi}=\int_{B}\delta C_{\xi}\;.\eeq
Imposing the linearized constrant $\delta C_{\xi}=0$, this simply becomes the first law of stretched future lightcones for higher derivative gravity.

\bibliography{referencesGR}

\end{document}